\documentclass[12pt]{iopartshmz}
\pdfoutput=1
\usepackage{iopams,bm}
\usepackage{graphicx}
\usepackage{cite}

\newcommand\kB{k_{\rm B}}
\newcommand\bra{\langle}   \newcommand\ket{\rangle}

\renewcommand\Re{{\rm Re}}
\renewcommand\Im{{\rm Im}}

\newcommand*\sub[1]{_{\rm{#1}}}
\newcommand*\braket[1]{\langle{#1}\rangle}

\begin{document}

\title[Quantum Violation of Fluctuation-Dissipation Theorem]{Quantum Violation of Fluctuation-Dissipation Theorem}

\author{Akira Shimizu and Kyota Fujikura}

\address{Department of Basic Science, The University of Tokyo\\ 
3-8-1 Komaba, Meguro-City, Tokyo 153-8902, Japan}
\ead{shmz@asone.c.u-tokyo.ac.jp}
\vspace{10pt}
\begin{indented}
\item[](J. Stat. Mech. (2017) 024004; Published 9 February 2017)
\end{indented}

\begin{abstract}
We study quantum measurements of temporal equilibrium fluctuations in macroscopic quantum systems.
It is shown that 
the fluctuation-dissipation theorem,
as a relation between observed quantities, 
is partially violated in quantum systems, 
even if measurements are made 
in an ideal way that emulates classical ideal measurements
as closely as possible.
This is a genuine quantum effect that survives on a macroscopic scale.
We also show that 
the state realized during measurements of 
temporal equilibrium fluctuations
is a `squeezed equilibrium state,'
which is macroscopically identical to the pre-measurement equilibrium state
but is squeezed by the measurement.
It is a time-evolving state, 
in which macrovariables fluctuate and relax.
We also explain some of subtle but important points,
careless treatments of which often lead to unphysical results, 
of the linear response theory.
\end{abstract}

%
%
%
%
%

\section{Introduction}\label{sec:intro}

The fluctuation-dissipation theorem (FDT) is
widely regarded as a universal 
relation between linear response functions and 
temporal equilibrium fluctuations 
which are expressed by time correlations
\cite{einstein,Johnson,Nyquist,Takahashi,Green,CW,Nakano,Kubo,Kubo2}.
It states
\begin{eqnarray}
\mbox{linear response function}
& \ = \
\beta \times 
\mbox{temporal equilibrium fluctuation},
\label{fdt1}
\\
& \ = \
\beta \times
\mbox{time correlation in equilibrium},
\label{fdt2}
\end{eqnarray}
where $\beta$ denotes the inverse temperature, $1/T$ (we take $\kB=1$).
There is another FDT, sometimes called the FDT of the second kind \cite{KTH}, 
which states that 
for the Langevin equation the noise correlation is related 
to the linear response function.
We here consider the above FDT, sometimes called 
the FDT of the first kind \cite{KTH},
which can be tested directly by experiments.

The FDT resembles the `fluctuation-response relation' in equilibrium statistical 
mechanics, 
which relates thermodynamic responses (such as the specific heat
and static magnetic susceptibility) to ensemble fluctuations
(such as the variance of the energy in the Gibbs ensemble).
However, the response functions in the FDT are nonequilibrium (dissipative) 
properties, whereas the thermodynamic responses 
in the fluctuation-response relation are equilibrium properties.
Because of this crucial difference, 
the former is related to the temporal fluctuations, which are
essentially time correlations,
whereas the latter is related to the ensemble fluctuations, 
which are essentially spatial correlations 
in the Gibbs ensemble.\footnote{
The term `equilibrium fluctuation' is often used 
both for the temporal fluctuations and 
for the ensemble fluctuation.
We avoid such a confusing terminology,
except when confusion is unlikely.
}
Hence, 
the FDT is considered as a fundamental relation in 
{\em nonequilibrium} statistical mechanics \cite{KTH}.
Although some parts of response functions, such as 
the imaginary part of the conductivity, 
are not directly related to dissipation,
the term FDT is used generally 
for relations \eref{fdt1} and \eref{fdt2} \cite{Kubo2}.



Experimentally, 
after Johnson discovered the FDT for the first time \cite{Johnson}, 
later papers did not seem to claim experimental evidence for 
the FDT (because Johnson already discovered it),
but rather they {\em utilized} the FDT to study other subjects.
For example, in 
experiments on the amplitude squeezing of light \cite{SQlaser,SQled},
the FDT played the central role as pointed out in \cite{OB}.
Another example is the FDT in mechanical experiments
(see, e.g., \cite{Saulson} and references cited therein).
The consistency of these results also evidenced the FDT.
However, most of these experimental evidences,  
including Johnson \cite{Johnson}, 
are limited to the real part of the {\em symmetric part} 
(as defined in \ref{sec:sandas}) of the response function, such as the real part of the diagonal conductivity $\Re \, \sigma_{xx}$, 
in the so-called {\em classical regime} $\hbar \omega \ll k_B T$.
Thus a question arises: Does the FDT really hold in other cases? That is, is the FDT really a universal relation?


We here show that the answer is no.
As a relation between observed quantities,
the FDT holds {\em only} in the above limited case. 
For example, it is violated for the real antisymmetric part of 
the admittance (which is the Fourier transform of response function) 
{\em even in the classical regime.}  A typical example is the Hall conductivity $\sigma_{xy}$ at $\omega=0$. 
The violation is a genuine quantum effect that appears on the macroscopic scale.
Even for the real symmetric part, 
for which the FDT holds in the classical regime, 
the FDT is violated in the non-classical regime $\hbar \omega \gtrsim k_B T$.

We also obtain explicitly the state 
that is realized during measurements of 
temporal equilibrium fluctuations.
It is a `squeezed equilibrium state,'
which is macroscopically identical to the pre-measurement equilibrium state
but is squeezed by the measurement.
It is a time-evolving state, 
in which macrovariables fluctuate and relax.
Such an interesting state should be 
realized during quasiclassical measurements of 
temporal equilibrium fluctuations.

Since the main points of the theory were already described in \cite{FS2016},
we emphasize physical aspects in this paper.
For clarity, 
we also explain some of subtle but important points
of the linear response theory,
careless treatments of which often lead to unphysical results
that are found in the literature.

The paper is organized as follows.
In \sref{sec:wrong},
we explain 
the problems in the previous derivations of the FDT for quantum systems.
We then describe the assumptions of the present analysis in 
\sref{sec:assumptions}.
In \sref{sec:mtc},
using modern theory of quantum measurements,
we analyze the process of measuring the time correlation.
We then show the violation of the FDT in \sref{sec:violation}.
These results, and related works, are discussed in \sref{sec:discussion}.
The paper is summarized in \sref{sec:summary}.

\section{What's wrong with derivations of FDT for quantum systems}
\label{sec:wrong}

\subsection{Classical systems}

For a classical system, 
Einstein discovered theoretically the first example of 
the FDT  \cite{einstein},
by assuming a stochastic model.
Another important example was discovered experimentally by Johnson \cite{Johnson}.
Nyquist presented an elegant theory that derived Johnson's results \cite{Nyquist}.
He first established the universality of the FDT from the second law of thermodynamics, and then
derived the FDT for classical electric circuits.
Furthermore, he introduced a quantum effect intuitively into his result.

In these works, 
Nyquist utilized a macroscopic theory (thermodynamics and circuit theory)
and Einstein utilized 
a mesoscopic theory (stochastic model).
Green also developed 
a mesoscopic theory \cite{Green}.
The microscopic derivation,
i.e., derivation from Newtonian mechanics of point particles, 
of the FDT
was given by Takahashi for general classical systems \cite{Takahashi}.

\subsection{Quantum systems}

The microscopic derivation by Takahashi \cite{Takahashi}
of the FDT for classical systems 
appeared generalizable to quantum systems
because of the similarity between classical and quantum mechanics.
However, he hesitated such generalization
because disturbances (backactions) caused by quantum measurements should be considered seriously.  
He stated
``This, however, introduces a rather profound difficulty, 
which originates in the very nature 
of quantum mechanical observation, 
that every observation disturbs the system" \cite{Takahashi}.

\begin{figure}[tp]
\begin{center}
\includegraphics[width=\textwidth]{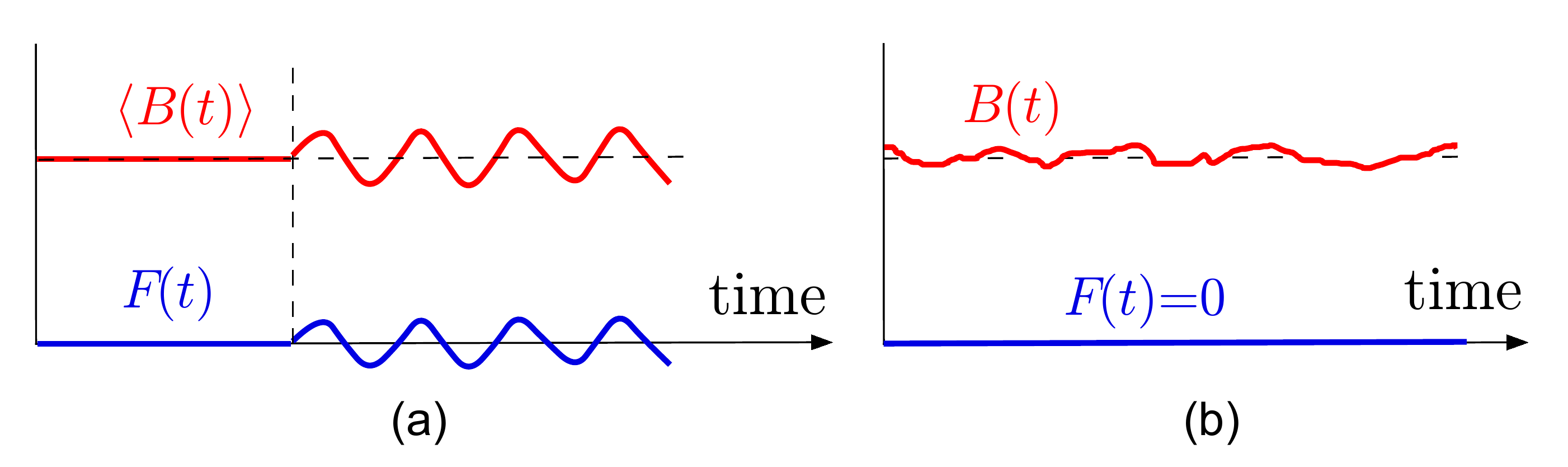}
\end{center}
\caption{An external field $F(t)$ and 
the time evolution of an additive observable $B$
when (a) the response function is measured, 
and (b) the temporal equilibrium fluctuation is measured.
}
\label{meas_resp_fluc}
\end{figure}

His concern may be understood by considering how the 
response functions and fluctuations are measured.
When measuring a response function, one applies an external field $F(t)$ 
to the system and measure 
the time variation of a macrovariable, 
say $B$, as shown in \Fref{meas_resp_fluc}(a).
When measuring the temporal equilibrium fluctuation of $B$, one 
takes $F(t)=0$ and measures the variation of $B$, 
as shown in \Fref{meas_resp_fluc}(b).
In both cases, one usually performs  
multi-time (or continuous) measurements 
to obtain values of $B$ at various times.
In the case of two-time measurements, for example, 
one measures $B$ at $t=t_0$ and {\em subsequently} at $t=t_1$.
Consequently, 
disturbance by the first 
measurement at $t=t_0$
affects the result of the second 
measurement at $t=t_1$.
Such a process is described not by the unitary time evolution (i.e., the Schr\"{o}dinger or von Neumann equation) but by a non-unitary evolution 
\cite{Glauber,Gardiner,MW,WM,KSreview}.

However, Callen and Welton \cite{CW}, Nakano \cite{Nakano}, and 
Kubo \cite{Kubo} 
studied the FDT for quantum systems assuming a unitary 
time evolution, completely neglecting the disturbances by measurements.
Actually, for the temporal equilibrium fluctuation 
Callen and Welton, Nakano and Kubo, 
and Nyquist (who introduced a quantum effect intuitively) \cite{Nyquist}
claimed different time correlations, 
which agree with each other only for classical systems.
(See, e.g., \ref{sec:example} and 
footnote \ref{fn:agreement} in \ref{sec:vio_high_f}.)
Among them, the most widely used for quantum systems seems 
Kubo's result,
which for the transverse 
electrical conductivity agrees with the previous work by Nakano \cite{Nakano}.
We therefore consider 
the `Kubo formula.'
It states (see \ref{sec:kuboformula} for details)
\begin{eqnarray}
\mbox{linear response function}
& \ = \
\beta \times
\mbox{canonical time correlation},
\label{kuboformula}\end{eqnarray}
where the {\em canonical time correlation} 
is defined by
\begin{equation}
\bra \hat{X}; \hat{Y}(t) \ket\sub{eq}
\equiv
{1 \over \beta} \int_0^\beta 
\bra 
e^{\lambda \hat{H}} \hat{X} e^{- \lambda \hat{H}} \hat{Y}(t)
\ket\sub{eq}
d \lambda
\label{canTC}
\end{equation}
for two Heisenberg operators $\hat{X} = \hat{X}(0)$ and 
$\hat{Y}(t) = e^{i \hat{H} t/\hbar} \hat{Y} e^{-i \hat{H} t/\hbar}$.
Here, 
$\hat{H}$ is the Hamiltonian, and 
$\braket{\ \cdot\ }\sub{eq}$ denotes 
the equilibrium expectation value. 

\subsection{What are the problems}

For classical systems, the FDT is considered to hold 
not just as a formal relation but as a relation between 
observed quantities.
This point is very important because the left- and right-hand sides 
of the FDT
correspond to quite different experiments, as shown in \Fref{meas_resp_fluc}.
This is why the FDT is very significant. 
For example, 
by measuring only the response function
one can deduce what will be observed when fluctuation is measured, 
and vice versa.

For quantum systems, however, this point has been unclear because,
as mentioned above, 
the previous derivations \cite{CW,Nakano,Kubo} neglected disturbances 
by measurements which are inevitable in quantum systems.
Thus, the question is: Does 
the FDT hold in quantum systems as relations between observed quantities?

\begin{figure}[tp]
\begin{center}
\includegraphics[width=\textwidth]{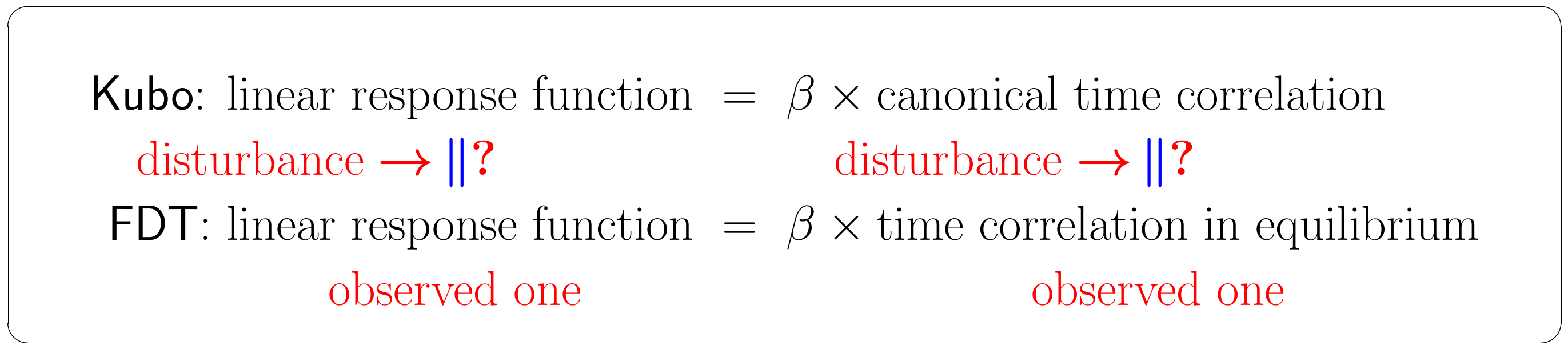}
\end{center}
\caption{Comparison of the Kubo formula and the FDT as 
a relation between observed quantities.
The question is whether each hand side of the two relations
agrees with each other in the presence of disturbances by quantum 
measurements.
}
\label{diagram1}
\end{figure}

By comparing \eref{fdt2} with \eref{kuboformula}, 
we can see what are the problems, as shown in \Fref{diagram1}.
One is whether the linear response function by the Kubo formula agrees with the observed one. 
The other is whether the canonical time correlation agrees with the observed time correlation. 
These problems have been left unsolved because at the time of the above pioneering works neither quantum measurement theory 
\cite{Glauber,Gardiner,MW,WM,KSreview} 
nor theory of macroscopic quantum systems 
\cite{goderis,matsui,Mthesis,Lieb,Nachtergaele,LRB_cont}
was developed enough. Fortunately, both these theories have been developed greatly in the last few decades.  
This enabled us to solve the problems. 

\begin{figure}[tp]
\begin{center}
\includegraphics[width=\textwidth]{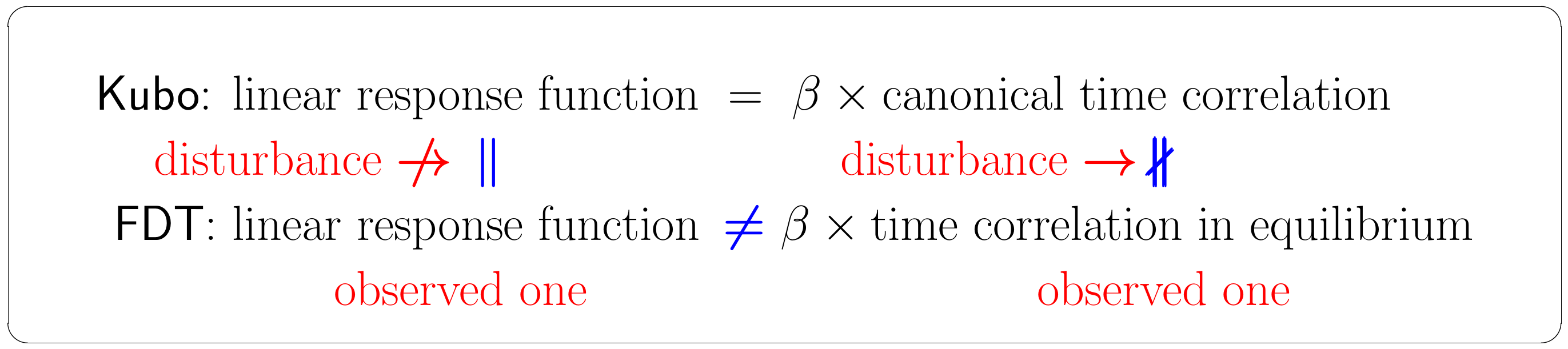}
\end{center}
\caption{Our result for the question raised in \Fref{diagram1},
and its consequence that the FDT is violated as a relation between observed quantities.
}
\label{diagram2}
\end{figure}

One might wonder if macrovariables can be affected considerably by quantum disturbances.
Our answer is no, when response is measured.
Hence, the Kubo formula may be correct {\em as a recipe} 
to calculate response functions.
However, the answer is yes, when fluctuation is measured. 
That is, the canonical time correlation does not agree with the observed time correlation.
Therefore, \Fref{diagram1}
is updated as \Fref{diagram2}.
Since the canonical time correlation does not agree with the observed time correlation, 
the FDT is violated as a relation between observed quantities.
We shall present how we have derived these results. 

\section{Assumptions}\label{sec:assumptions}

\subsection{Assumptions on the system and its equilibrium states}\label{sec:assmp.eq}

We consider a $d$-dimensional macroscopic system ($d=1, 2, 3, \cdots$) of size $N$ (which is, e.g., the number of spins).
As the pre-measurement state
we take an equilibrium state of temperature $T$ ($=1/\beta$),
which is assumed to be uniform macroscopically.\footnote{
When a phase separation occurs, 
apply the following results to each unifrom phase.
}
As the microscopic representation of the equilibrium state, we employ
the `canonical thermal pure quantum state' $|\beta \ket$,  
introduced and studied in \cite{SS2013}.  
It is a pure quantum state that shares all macroscopic properties with the canonical Gibbs state \cite{SS2013,vonN,SugitaJ,Popescu,Goldstein,SugitaE,Reimann2007,SS2012,Kindai2013,HSS2014}. 
 (This is true for any systems including 
integrable systems and random systems 
exhibiting many-body localization, 
where the eigenstate thermalization hypothesis \cite{vonN,ETH1,ETH2,ETH3} 
is violated.)
Hence, the equilibrium expectation value is 
obtained simply as the quantum-mechanical expectation value, 
$\braket{ \ \cdot \  }\sub{eq} = \braket{ \beta | \cdot | \beta }$.
Although the use of $|\beta \ket$ simplifies equations,
the reader, if prefers, may thoroughly 
replace it with the canonical Gibbs state $\hat{\rho}\sub{eq}$ 
appropriately in the following results.
[For example, 
formula \eref{eq:beta_adot} for the post-measurement state 
should be replaced with that for a mixed state.]

We assume that the correlation between any local observables 
at two points $\bm{r}$ and $\bm{r}'$ 
decays faster than 
$1/|\bm{r} - \bm{r}'|^{d+\epsilon}$, 
where $\epsilon$ is a positive constant.
This assumption is believed to hold generally, except at critical points.
Consequently, 
for all additive observable $\hat{A}$,
its ensemble fluctuation \cite{LL,SM2002,SM2005}
\begin{equation}
\delta A\sub{eq}
\equiv \sqrt{ \braket{ (\Delta \hat{A})^2  }\sub{eq}}= O(\sqrt{N}).
\end{equation}
Here, 
$\Delta \hat{A} \equiv \hat{A} - \braket{\hat{A}}\sub{eq}$.
[Throughout this paper, $\Delta$ denotes 
deviation from the equilibrium value.]
We also make additional reasonable assumptions 
(see Supplemental Material of \cite{FS2016} for details).
Then, the quantum central limit theorem (QCLT) holds
\cite{goderis,matsui,FS2016,Mthesis}, 
from which we can draw 
the universal results 
presented in \sref{sec:mtc} and \sref{sec:violation}.

We assume that an additive observable 
is the sum of the {\em same} local observable over the whole system,
where `local observable' means an observable on a finite region 
whose size is independent of $N$.
For example, the staggered magnetization and
the total current 
are additive observables.

\subsection{Assumptions on measurements}\label{sec:assmp.ms}

Suppose that one tries to determine
temporal equilibrium fluctuations
by measuring time correlations.
If a violent detector were used, 
it would destroy completely the state by the first measurement,
and consequently a meaningless result would be obtained for the 
second measurement.
As a result, a wrong result would be obtained for the correlation, 
and obviously the FDT would look violated.
Therefore, 
certain ``ideal" detectors should be used to measure the time correlation correctly. 


In classical systems, an ideal detector is trivially defined as a detector that does not disturb the state at all. 
In quantum systems, however, such a detector is impossible because of 
the uncertainty relation \cite{AG}.
Hence, to inspect the validity of the FDT in quantum systems, 
the best possible way is to use a detector that emulates the classical ideal one as closely as possible. 
We call such a detector {\em quasiclassical}.

We note that a quasiclassical measurement
should have a moderate magnitude of error.
For measuring the temporal equilibrium fluctuation, 
which is of the same order as the ensemble fluctuation $\delta A\sub{eq}$, 
the measurement error $\delta A\sub{err}$ should be smaller,
$\delta A\sub{err} < \delta A\sub{eq}$.
On the other hand, 
$\delta A\sub{err}$ should not be too small because 
the disturbance increases with decreasing $\delta A\sub{err}$, 
according to the uncertainty relation 
between error and disturbance \cite{AG}.
We therefore require
\begin{equation}
\delta A\sub{err} = \varepsilon \delta A\sub{eq},
\label{Aerr=eAeq}
\end{equation}
where $\varepsilon$ is a small positive number independent of $N$.
[Actually, the following results hold also for larger $\varepsilon$
(if it is independent of $N$), which occurs, e.g.,
when interaction with the measuring apparatus is weak \cite{KSreview,AG}.]
Since $\delta A\sub{eq} = O(\sqrt{N})$ as mentioned above, this means
$\delta A\sub{err} = O(\sqrt{N})$.
Therefore, to formulate measurements of temporal equilibrium fluctuations, 
we should scale additive operators as
\begin{equation}
\hat{a} = \hat{A}/\sqrt{N}.
\end{equation}
We shall use such scaled operators. 
[Otherwise, some of the following equations would diverge in the 
thermodynamic limit.]

The general framework of quantum measurement 
\cite{KSreview,NC,wiseman2010}
can be adapted to our problem as follows.
Let $|\psi \ket$ be the pre-measurement state
that is uniform macroscopically, 
such as $| \beta \ket$. 
Suppose that $\hat{A}$ is measured.
We denote the outcome of the measurement by 
$A_\bullet$, 
which is a real valued variable.
Since a quasiclassical measurement has 
a non-vanishing error, 
$A_\bullet$ does not necessarily agree with one of eigenvalues of $\hat{A}$. 
Moreover, $a_\bullet \equiv A_\bullet/\sqrt{N}$ can be regarded as a continuous variable,
even when $\hat{A}$ has a discrete spectrum (whose spacing 
is $O(1)$ because $\hat{A}$ is an additive observable).
The probability density
of getting $a_\bullet$ is given by
the {\em probability operator} $\hat{E}_{a_\bullet}$,
which is a Hermitian positive semidefinite operator such that 
the integral over the outcome is the identity operator,
as
\begin{equation}
p(a_\bullet) 
= 
\bra \psi | \hat{E}_{a_\bullet} | \psi \ket.
\end{equation}
The probability operator can be represented as the product of 
{\em measurement operator} $\hat{M}_{a_\bullet}$ as
\begin{equation}
\hat{E}_{a_\bullet}
= \hat{M}_{a_\bullet}^\dagger \hat{M}_{a_\bullet}.
\end{equation}
The post-measurement state is given by the measurement operator as
\begin{equation}
\mbox{post-measurement state}
=
\sqrt{1/p(a_\bullet)} \
\hat{M}_{a_\bullet} |\psi \ket,
\label{mt_post}
\end{equation}
where the prefactor $\sqrt{1/p(a_\bullet)}$ is just a normalization constant.

Using this general framework, we precisely define 
{\em quasiclassical measurement} of an additive observable as follows.
\begin{itemize}
\item[(i)] 
It is unbiased, i.e., 
\begin{equation}
\overline{a_\bullet} = \braket{\hat{a}}\sub{eq}, 
\end{equation}
where $\overline{\cdots}$ denotes the average over many runs of experiments.
That is, by averaging the outcomes over many runs of experiments,
one obtains the correct expectation value.
(Otherwise, the FDT would look more violated.)

\item[(ii)] 
$\hat{E}_{a_\bullet}$ scales in such a way that the probability distribution of $A_\bullet$ 
in $| \beta \ket$ scales as $\sqrt{N}$ 
apart from the uniform shift associated with 
$\braket{\hat{A}}\sub{eq} \propto N$,
i.e., 
$p\sub{shifted}(\Delta a_\bullet) 
\equiv p(\Delta a_\bullet + \braket{\hat{a}}\sub{eq})$ 
converges as $N \to \infty$.
This yields, e.g., 
$\delta a\sub{err} = O(1)$, 
i.e., 
$\delta A\sub{err} = O(\sqrt{N})$, 
as required.

\item[(iii)] 
$\hat{M}_{a_\bullet}$ is minimally disturbing \cite{wiseman2010}
among many possible 
measurement operators that give the same $\hat{E}_{a_\bullet}$.
That is \cite{wiseman2010}, 
\begin{equation}
\hat{M}_{a_\bullet} = \sqrt{\hat{E}_{a_\bullet}}.
\end{equation}

\item[(iv)] 
It is homogeneous, i.e., 
$\hat{E}_{a_\bullet}$ depends on $\hat{a}$ and $a_\bullet$ only through 
$\hat{a} - a_\bullet$.
This yields, e.g., a reasonable property that 
$\delta a\sub{err} =$ independent of $a_\bullet$.
(Otherwise, analysis of experimental results would be complicated.)

From (i)-(iv), we find 
\begin{equation}
\hat{M}_{a_\bullet} = f(\hat{a}-a_\bullet),
\end{equation}
where $f(x) \geq 0$.

\item[(v)]
$f(x)$ behaves well enough, e.g., it vanishes quickly as $|x| \to \infty$.
[Detailed conditions are described in Supplemental Material of \cite{FS2016}.]

\end{itemize}
Roughly speaking, 
we say measurement is quasiclassical if
it is minimally-disturbing, homogeneous, and unbiased, with a moderate magnitudes of error (which is small enough to measure fluctuations but not too small in order to avoid strong disturbances). 
Concretely, the measurement error
$\delta A\sub{err} = \sqrt{N} \delta a\sub{err}$
is determined by $f(x)$ as
\begin{equation}
\delta a\sub{err}^2 = \int x^2 |f(x)|^2 dx = O(1),
\end{equation}
in consistency with \eref{Aerr=eAeq}.
%
%

A typical example is the case of 
the {\em Gaussian measurement operator}, 
for which 
\begin{eqnarray}
f(x) 
&= 
\frac{1}{(2\pi w^2)^{1/4}}\exp \left( - \frac{x^2}{4w^2} \right),
\quad
w=O(1)>0,
\label{Gaussianf}
\\
\hat{M}_{a_\bullet} 
&= f(\hat{a}-a_\bullet)
=
\frac{1}{(2\pi w^2)^{1/4}}
\exp \left[ - \frac{(\hat{a}-a_\bullet)^2}{4w^2} \right],
\\
\delta a\sub{err}^2
&= 
w^2 = O(1)
\qquad (\mbox{i.e., } \delta A\sub{err} = w \sqrt{N} = O(\sqrt{N})).
\end{eqnarray}

\section{Measurement of time correlation}\label{sec:mtc}

In this section, we study what is obtained when 
temporal equilibrium fluctuation is measured.
The reader, if not interested in the measurement process, may 
jump to the last paragraphs of \ref{sec:1stm} and
\ref{sec:2ndm}, where the main results of \sref{sec:mtc} are summarized.

\subsection{Measurement process}\label{sec:mpr}

\begin{figure}[tp]
\begin{center}
\includegraphics[width=\textwidth]{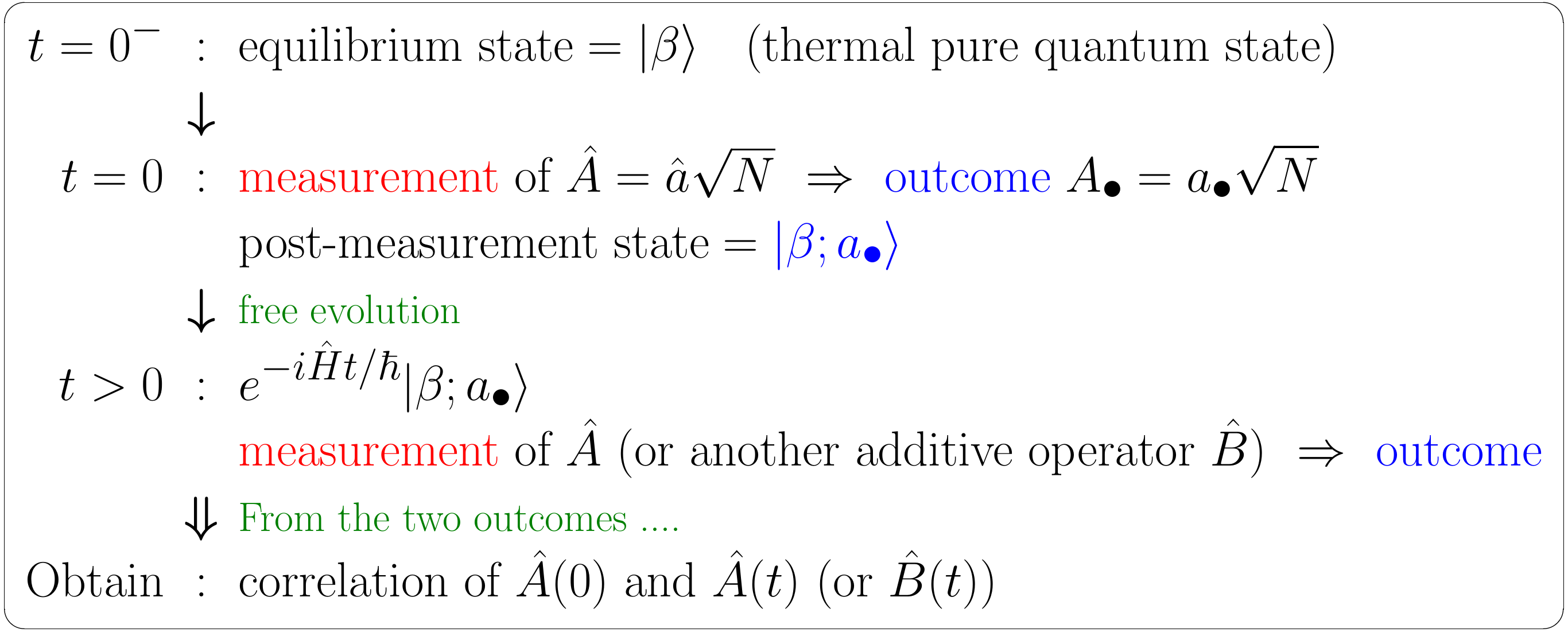}
\end{center}
\caption{A process of measurement of 
the time correlation of an additive observable at $t=0$, $\hat{A}(0)$, 
and that (or another additive operator $\hat{B}$)
at $t >0$, $\hat{A}(t)$.}
\label{diagram3}
\end{figure}

In \Fref{diagram3} we show a process of measurement of 
the time correlation of an additive observable at $t=0$, $\hat{A}(0)$, 
and that (or another additive operator $\hat{B}$)
at $t >0$, $\hat{A}(t)$.
The state just before the first measurement is an equilibrium 
state, which is represented by the canonical thermal pure quantum state 
$| \beta \ket$ \cite{SS2013}
as mentioned in \ref{sec:assmp.eq}.
At $t=0$, 
the first measurement of $\hat{A} = \hat{a} \sqrt{N}$
is made, 
and the outcome $A_\bullet = a_\bullet \sqrt{N}$ is obtained.
The post-measurement state is denoted by $| \beta; a_\bullet \ket$.
It evolves freely as 
$e^{-i \hat{H} t/\hbar} | \beta; a_\bullet \ket$,
until the second measurement of $\hat{A}$ 
(or $\hat{B}$) is made at $t >0$,
and the outcome of this measurement is obtained.
From the outcomes of the first and the second measurements, 
one obtains the correlation of $\hat{A}(0)$ and $\hat{A}(t)$ (or $\hat{B}(t)$).

In this process, 
the first measurement should be quasiclassical in order to minimize 
the disturbance, which affects the result of the second measurement.
On the other hand, 
the second measurement is not required to be quasiclassical
(e.g., one may use the projection measurement)
because its post-measurement state will not be measured and hence
the disturbance is irrelevant.

We shall analyze this process step by step.
Since our results are derived 
using the QCLT 
\cite{goderis,matsui,FS2016,Mthesis},
they hold universally, irrespective of details of the system,
provided that the system satisfies the assumptions in \ref{sec:assmp.eq}.
%
Although the main results hold in the thermodynamic limit, 
we do not write the limit symbol explicitly
except when we want to stress it.

While this process assumes that measurements are performed twice
in each run of experiments,
more general processes, in which measurement is performed three or more times in each run, 
are also analyzed in \cite{FS2016}.

\subsection{First measurement and its disturbances}
\label{sec:1stm}

For the first measurement at $t=0$,
we can calculate using the QCLT 
\cite{goderis,matsui,FS2016,Mthesis}
the probability density of getting a particular value of 
the outcome $a_\bullet$.
For the Gaussian $f$ of \eref{Gaussianf}, for example, 
it is calculated as
\begin{equation}
\fl
p(a_\bullet)
=
{ 
1
\over
[2 \pi (\delta a\sub{eq}^2 + \delta a\sub{err}^2)]^{1/2}
}
\exp \left[ 
-
{ 
1
\over
2 (\delta a\sub{eq}^2 + \delta a\sub{err}^2)
}
(\Delta a_\bullet)^2
\right]
\qquad \mbox{(for Gaussian $f$)},
\label{p(adot)}
\end{equation}
where 
$\delta a\sub{eq} \equiv \delta A\sub{eq}/\sqrt{N}$
and 
\begin{equation}
\Delta a_\bullet \equiv a_\bullet -\braket{\hat{a}}\sub{eq}.
\end{equation}
It is seen that 
the width of the distribution of $a_\bullet$ 
is larger than the actual width $\delta a\sub{eq}$ of $\hat{a}$,
because of the measurement error $\delta a\sub{err}$.
For general $f$, we obtain a similar result:  
\begin{equation}
p(a_\bullet) 
= 
\int 
{
\left|f(x)\right|^2
\over (2\pi \delta a\sub{eq}^2)^{1/2}
}
\exp\left[
- 
{
(x+\Delta a_\bullet)^2
\over 
2\delta a\sub{eq}^2
}
\right]
dx.
\label{eq:prob_A_by_qclt}
\end{equation}
It is a convolution of 
the distribution of $\hat{a}$ in 
$| \beta \ket$,
which is Gaussian according to the QCLT, 
and the shape $|f(x)|^2$ of 
the measurement operator.

\begin{figure}[tp]
\begin{center}
\includegraphics[width=0.8\textwidth]{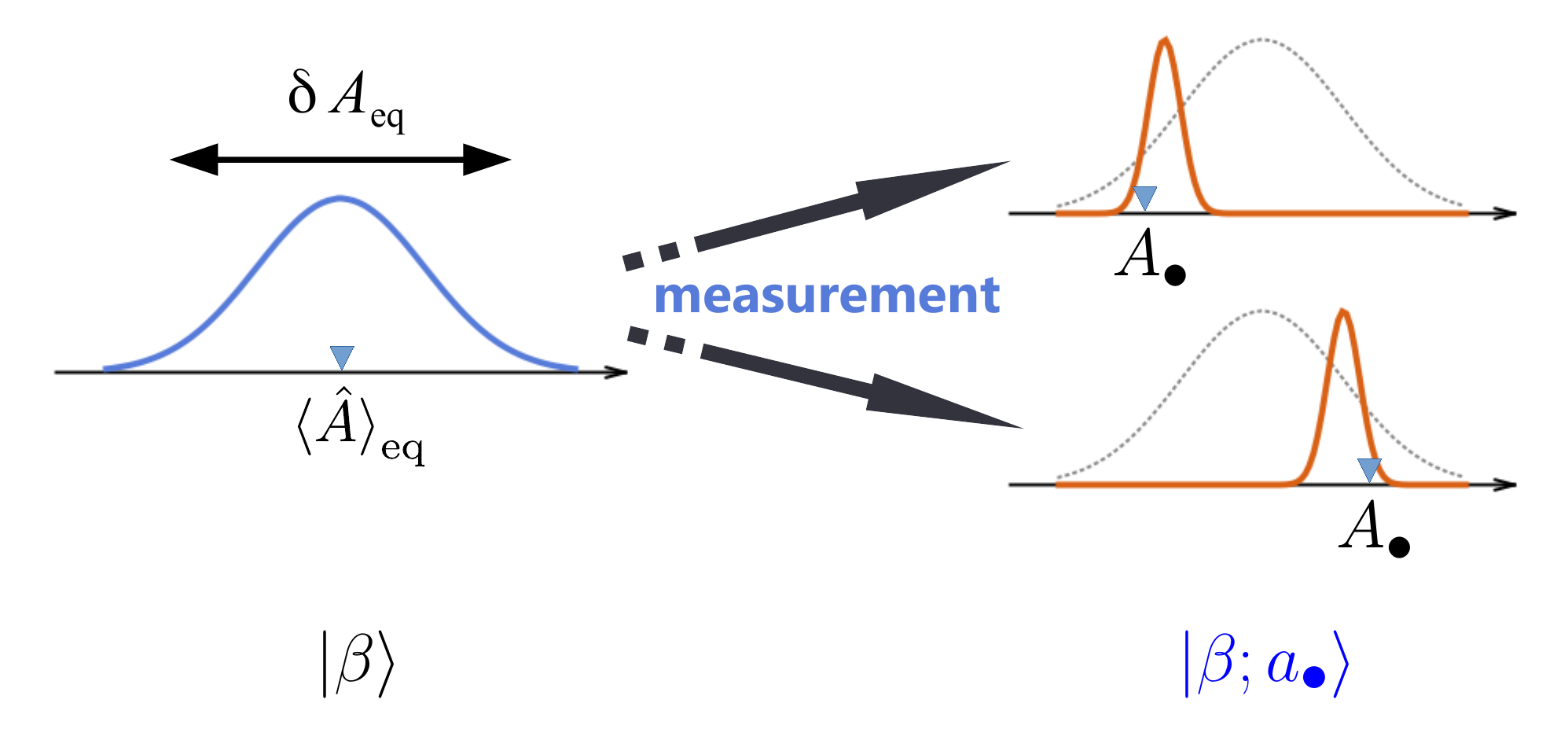}
\end{center}
\caption{Schematic plots of the distribution of an additive observable $A$ 
in the pre-measurement state $| \beta \ket$ (left)
and 
in the post-measurement state $| \beta; a_\bullet \ket$ (right).
}
\label{post_m_state_a}
\end{figure}

The post-measurement state $| \beta; a_\bullet \ket$ is given,
according to \eref{mt_post}, by
\begin{equation}
| \beta; a_\bullet \ket
=
{1 \over \sqrt{p(a_\bullet)}}
f( \hat{a} - a_\bullet ) | \beta \ket.
\label{eq:beta_adot}
\end{equation}
To investigate its properties, we calculate
the expectation value and variance of $\hat{a}$ in this state.
We denote 
$\braket{\ \cdot \ }_{a_\bullet} \equiv 
\bra \beta; a_\bullet | \cdot | \beta; a_\bullet \ket$.
For the Gaussian $f$, we find
\begin{equation}
\braket{\Delta \hat{a}}_{a_\bullet}
\equiv
\braket{\hat{a}}_{a_\bullet}
-
\braket{\hat{a}}\sub{eq}
%
%
=
{ 
\delta a\sub{eq}^2
\over
\delta a\sub{eq}^2 + \delta a\sub{err}^2
}
\Delta a_\bullet
\qquad \mbox{(for Gaussian $f$)},
\label{Deltaaad}
\end{equation}
which shows that 
$\hat{a}$ is shifted toward the outcome
as a result of the ``collapse of the wavefunction."
%
For the variance, we find
\begin{equation}
\langle (\hat{a} - \langle \hat{a} \rangle_{a_\bullet})^2 \rangle_{a_\bullet}
=
\left[ 
1
\ - \
{
{\delta a\sub{eq}^2}
\over
\delta a\sub{eq}^2 + {\delta a\sub{err}^2}
}
\right]
{\delta a\sub{eq}^2}
\qquad \mbox{(for Gaussian $f$)},
%
\end{equation}
which shows that the state is `squeezed' ( i.e., 
the variance is reduced) along $\hat{a}$.
These results are summarized schematically in \Fref{post_m_state_a}.

For another additive operator $\hat{B} = \hat{b} \sqrt{N}$,
we have, for the Gaussian $f$,
\begin{equation}
\braket{\Delta \hat{b}}_{a_\bullet}
\equiv
\braket{\hat{b}}_{a_\bullet}
-
\braket{\hat{b}}\sub{eq}
=
{
\langle 
\frac{1}{2} \{ \Delta \hat{a}, \Delta \hat{b} \}
\rangle\sub{eq}
\over
\delta a\sub{eq}^2 + \delta a\sub{err}^2
}
\Delta a_\bullet
\qquad \mbox{(for Gaussian $f$)},
\end{equation}
where
$ 
\frac{1}{2} \{ \hat{X}, \hat{Y} \}
\equiv 
\frac{1}{2} \left( \hat{X}\hat{Y} + \hat{Y}\hat{X} \right)
$
is the anticommutator.
It shows that $\hat{b}$ is also shifted 
as a result of the measurement, 
and the sign and magnitude of the shift depend on 
those of $\langle \frac{1}{2} \{ \Delta \hat{a}, \Delta \hat{b} \} \rangle\sub{eq}$.
We also have
\begin{equation}
\fl
\langle (\hat{b} - \langle \hat{b} \rangle_{a_\bullet})^2 \rangle_{a_\bullet}
=
\delta b\sub{eq}^2
-
{
\langle 
\frac{1}{2} \{ \Delta \hat{a}, \Delta \hat{b} \}
\rangle\sub{eq}^2
\over
\delta a\sub{eq}^2 + \delta a\sub{err}^2
}
+
{
\braket{\mbox{$\frac{1}{2i}$} [ \hat{a}, \hat{b} ]}\sub{eq}^2
\over 
\delta a\sub{err}^2
}
\qquad \mbox{(for Gaussian $f$)}.
\label{varbad}
\end{equation}
The second term in the right-hand side 
represents the reduction of the variance by the measurement, while
the third term shows that the variance increases 
by the measurement if 
$[\hat{a}, \hat{b}]$ is non-negligible in $| \beta \ket$.
Since the magnitude of the variance is determined by these 
competing terms,  the variance of $\hat{b}$ 
is not necessarily reduced by
the measurement of $\hat{a}$ when $[ \hat{a}, \hat{b} ] \neq 0$.
For general $f$, we obtain similar results
(the cases of $t=0$ in \eref{eq:<B>} and in Eq.~(13) of \cite{FS2016}),
which depend on $f$.

Since the right-hand sides of \eref{Deltaaad}-\eref{varbad}
are $O(1)$, we find that 
{\em disturbances on additive operators $\hat{A}, \hat{B}, ...$ by quasiclassical measurements are $O(\sqrt{N})$},
which is of the same order as the equilibrium fluctuations.
This means that 
the post-measurement state 
$| \beta; a_\bullet \ket$ is 
{\em macroscopically identical} to the 
pre-measurement equilibrium state $| \beta \ket$,
although $| \beta; a_\bullet \ket$ is squeezed along $\hat{a}$.
Hence, 
$| \beta; a_\bullet \ket$ may be called 
a {\em squeezed equilibrium state}.
We stress that 
such a state is always realized after the above measurement process, which seems to be a typical and reasonable procedure for measuring fluctuation.

\subsection{Second measurement and obtained time correlation}
\label{sec:2ndm}

The post-measurement state evolves freely as 
$e^{-i \hat{H} t/\hbar} | \beta; a_\bullet \ket$,
until the second measurement 
is made at $t >0$.
When an additive observable $\hat{B} = \hat{b} \sqrt{N}$ is measured 
in this second measurement, 
its expectation value is calculated for the Gaussian $f$ as
\begin{eqnarray}
\braket{\Delta \hat{b}(t)}_{a_\bullet}
&\equiv
\braket{\hat{b}(t)}_{a_\bullet}
-
\braket{\hat{b}}\sub{eq}
\nonumber\\
&=
\Theta(t)
\langle 
\mbox{$\frac{1}{2}$} \{ \Delta \hat{a}, \Delta \hat{b}(t) \}
\rangle\sub{eq}
{
\Delta a_\bullet
\over
\delta a\sub{eq}^2 + \delta a\sub{err}^2
}
\qquad \mbox{(for Gaussian $f$)}.
\end{eqnarray}
Here, $\Theta(t)$ is the step function, which says simply that if this measurement is made before the first one
then the equilibrium value will be obtained 
(i.e., $\braket{\hat{b}(t)}_{a_\bullet} = \braket{\hat{b}}\sub{eq}$).
For general $f$, we obtain a similar result:  
\begin{equation}
\braket{\Delta \hat{b}(t)}_{a_\bullet}
=
-
\Theta(t)
\langle 
\mbox{$\frac{1}{2}$}
\{ \Delta \hat{a}, \Delta \hat{b}(t) \}
\rangle\sub{eq}
{d \ln p  \over d a_\bullet},
\label{eq:<B>}
\end{equation}
where $p$ is given by \eref{eq:prob_A_by_qclt}.
It is seen that the $t$ dependence of 
$\braket{\Delta \hat{b}(t)}_{a_\bullet}$ is governed by 
$\langle 
\mbox{$\frac{1}{2}$} \{ \Delta \hat{a}, \Delta \hat{b}(t) \}
\rangle\sub{eq}
$, 
i.e., by the {\em symmetrized time correlation}, 
which is defined 
(for general $\hat{X} = \hat{X}(0)$ and $\hat{Y}(t) =
e^{i \hat{H} t / \hbar} \hat{Y} e^{-i \hat{H} t / \hbar}$) 
by 
\begin{equation}
\langle
\mbox{$\frac{1}{2}$}
\{ \hat{X}, \hat{Y}(t) \}
\rangle\sub{eq}
\equiv
\langle 
\mbox{$\frac{1}{2}$}
(\hat{X} \hat{Y}(t) + \hat{Y}(t) \hat{X})
\rangle\sub{eq}.
\end{equation}

The correlation of $\hat{a}(0)$ and $\hat{b}(t)$ is obtained 
from the results of the two subsequent measurements as
$\Delta a_\bullet \braket{\Delta \hat{b}(t)}_{a_\bullet}$.
One is interested in its average 
over many runs of experiments 
(such average is denoted by the overline), 
\begin{equation}
\Xi_{ba}(t) 
\equiv
\overline{
\Delta a_\bullet
\braket{\Delta \hat{b}(t)}_{a_\bullet}
}
\qquad (t>0).
\end{equation}
It is calculated from \eref{eq:<B>} as
\begin{eqnarray}
\Xi_{ba}(t) 
&=
\int
\Delta a_\bullet
\braket{\Delta \hat{b}(t)}_{a_\bullet}
\ p(a_\bullet) d a_\bullet
\nonumber\\
&=
- 
\langle 
\mbox{$\frac{1}{2}$}
\{ \Delta \hat{a}, \Delta \hat{b}(t) \}
\rangle\sub{eq}
\int
\Delta a_\bullet
p'(a_\bullet)
d a_\bullet
\nonumber\\
&=
\langle 
\mbox{$\frac{1}{2}$}
\{ \Delta \hat{a}, \Delta \hat{b}(t) \}
\rangle\sub{eq}
\int
p(a_\bullet) 
d a_\bullet
\nonumber\\
&=
\langle 
\mbox{$\frac{1}{2}$}
\{ \Delta \hat{a}, \Delta \hat{b}(t) \}
\rangle\sub{eq}
\qquad (t>0)
\label{eq:corr.ab:t>0}
\end{eqnarray}
{\em for all $f$.}
[Actually, this derivation is rather naive.
A rigorous derivation is given in Supplemental Material of \cite{FS2016}.]
Note that the last line does not depend on $f$ at all, 
although both $\braket{\Delta \hat{b}(t)}_{a_\bullet}$ and
$p(a_\bullet)$ in the first line depend on the form of $f$.
That is, \eref{eq:corr.ab:t>0} is a universal result,
which holds for all quasiclassical measurements.

When $\hat{B}$ is measured at $t<0$ before $\hat{A}$ is measured at $t=0$, 
their correlation is given by 
$\langle \mbox{$\frac{1}{2}$} 
\{ \Delta \hat{b}(t), \Delta \hat{a} \}
\rangle\sub{eq}$.
Since this is identical to \eref{eq:corr.ab:t>0} except that $t<0$ here,
we can summarize these results as 
\begin{eqnarray}
\tilde{\Xi}_{ba}(t)
&\equiv
\mbox{correlation of $\hat{a}$ and $\hat{b}(t)$}
\nonumber\\
&=
\langle 
\mbox{$\frac{1}{2}$}
\{ \Delta \hat{a}, \Delta \hat{b}(t) \}
\rangle\sub{eq}
\label{eq:corr.ab}
\end{eqnarray}
{\em for all $t$ and $f$.}

We have thus obtained the following universal conclusion:
{\em When temporal equilibrium fluctuations of macrovariables are measured
in an ideal way that emulates classical ideal measurements
as closely as possible,
the symmetrized time correlation \eref{eq:corr.ab} is always obtained}, 
among many quantum correlations that reduce to the same classical correlation as $\hbar \to 0$.
In other words, if one employs the symmetrized time correlation
when quantizing a classical model,
the results for the time correlation will be free from disturbances 
provided that the measurements are quasiclassical. 


\section{Violation of FDT}\label{sec:violation}


\subsection{Kubo formula and its necessary conditions}
\label{sec:kuboformula}

Suppose that an external field $F(t)$ is applied to an equilibrium state.
The system is driven to a nonequilibrium state.
We are interested in the change of an additive observable $\hat{B}$
from its equilibrium value $\braket{\hat{B}}\sub{eq}$ 
to the nonequilibrium value $\braket{\hat{B}}_t$,
where $\braket{\ \cdot \ }_t$ denotes the expectation value 
in the presence of $F(t)$.
When $F(t)$ is small enough,
this change is related linearly to $F(t)$
as
\begin{equation}
{\braket{\hat{B}}_t \over N} - {\braket{\hat{B}}\sub{eq} \over N}
=
\int_{-\infty}^t \Phi_{ba}(t-t') F(t') dt'.
\label{LRR}
\end{equation}
This relation is called the linear-response relation, 
and the function $\Phi_{ba}(t)$ is called the 
{\em (linear) response function}.
Although this relation is sometimes written in terms of 
$\braket{\hat{B}}_t - \braket{\hat{B}}\sub{eq}$,
one should consider 
$\braket{\hat{B}}_t/N - \braket{\hat{B}}\sub{eq}/N$
to take the thermodynamic limit, which is necessary 
for the Kubo formula \eref{kubof}.
As in the case of a symmetry-breaking field in equilibrium statistical 
mechanics, we take $F(t)=O(1)$.
Hence, $\Phi_{ba} = O(1)$.

Assume that the interaction Hamiltonian between $F(t)$ and 
the system takes the following form,
\begin{equation}
\hat{H}\sub{ext}(t) = - F(t) \hat{C},
\label{eq:Hext}
\end{equation}
where $\hat{C}$ is an additive observable of the system.
Kubo \cite{Kubo} showed, using the first-order 
perturbation theory in powers of $F(t)$,  
that 
\begin{eqnarray}
\Phi_{ba}(t)
&= 
\Theta(t) \lim_{N \propto V \to \infty} 
\frac{1}{i \hbar} \braket{ [\hat{c}, \hat{b}(t)] }\sub{eq},
\label{kubofcomm}
\\
&=
\Theta(t) \lim_{N \propto V \to \infty} 
\beta \braket{ \Delta \hat{a} ; \Delta \hat{b}(t) }\sub{eq},
\label{kubof}
\end{eqnarray}
which is called the {\em Kubo formula}.
[We have rewritten his formula slightly using
$
\braket{\hat{a}}\sub{eq}
= \left. {d \over dt} \braket{\hat{c}(t)}\sub{eq} \right|_{t=0}
=0
$.]
Here, 
the step function $\Theta(t)$ represents the causality,
and
$\displaystyle \lim_{N \propto V \to \infty}$ denotes the thermodynamic limit
where $N$ and $V$ (and other extensive variables) 
go simultaneously to infinity while keeping their ratio(s) constant.
Although not taken in Kubo's paper \cite{Kubo}, 
{\em this limit is necessary for avoiding 
the quantum recurrence} \cite{Zubarev}.
[As noted in \ref{app:tdlkubo}, a special care is necessary 
when this limit and other limits are taken.]
Furthermore, 
\begin{equation}
\hat{a} \equiv \hat{A}/\sqrt{N}, 
\ \hat{b} \equiv \hat{B}/\sqrt{N},
\ \hat{c} \equiv \hat{C}/\sqrt{N}
\end{equation}
are scaled additive observables, 
\begin{equation}
\hat{A}
\equiv 
\left. {d \over dt} \hat{C}(t) \right|_{t=0}
= \frac{1}{i \hbar} [ \hat{C}, \hat{H} ]
\label{eq:A=dC}
\end{equation}
is the velocity of $\hat{C}$
(where $\hat{H}$ is the Hamiltonian in the absence of $F(t)$),
and $\braket{ \Delta \hat{a} ; \Delta \hat{b}(t) }\sub{eq}$ is the 
{\em canonical time correlation} defined by \eref{canTC}.
Although \eref{kubofcomm} is convenient for practical calculations,
\eref{kubof} is often more appropriate for studying fundamental problems.

Note that it is necessary 
for the applicability of the Kubo formula
to take $\hat{H}$
in such a way that 
the limits \eref{kubofcomm}
and \eref{kubof} converge.\footnote{
This is possible even if 
the perturbation series in powers of $F(t)$ does not converge.}
Furthermore, 
the condition\footnote{
The order of the two limits, 
$\displaystyle \lim_{t \to \infty}$ 
and $\displaystyle \lim_{N \propto V \to \infty}$,
cannot be inverted because of the quantum recurrence.
See \ref{app:tdlkubo} for a related discussion.
}
\begin{equation}
\lim_{t \to \infty} 
\lim_{N \propto V \to \infty} 
\braket{ \Delta \hat{a} ; \Delta \hat{b}(t) }\sub{eq}
=0
\label{remote_for_kubof}
\end{equation}
is necessary 
because otherwise it would give an unphysical result
that $F(t)$ at a remote past would affect the present
nonequilibrium state.
The above condition implies, e.g., that 
\begin{equation}
[ \hat{A}, \hat{H} ] \neq 0 
\mbox{ and }
[ \hat{B}, \hat{H} ] \neq 0.
\label{noncomm_for_kubof}
\end{equation}
%
A related necessary condition is
\begin{equation}
\lim_{\epsilon \searrow 0} 
\lim_{N \propto V \to \infty}
\epsilon \int_0^{\infty} 
\braket{ \hat{C}/N ; \hat{B}(t)/N }\sub{eq}
e^{- \epsilon t} dt
=
\lim_{N \propto V \to \infty}
\braket{ \hat{C}/N }\sub{eq} \braket{ \hat{B}/N }\sub{eq},
\label{mixing_for_kubof}
\end{equation}
%
%
where the corresponding condition in \cite{Kubo} has been 
made precise 
using the results of \ref{app:tdlkubo}.
The conditions \eref{remote_for_kubof} and \eref{mixing_for_kubof}
resemble the ``mixing property" in classical dynamics.

From these conditions,
in general, 
it is not justified to apply the Kubo formula to integrable systems 
that have many additive conserved observables,
though such erroneous application is often found in the literature 
(see \cite{SK} for detailed discussions and a concrete example).
We henceforth {\em assume that the above necessary conditions 
are all satisfied.}

\subsection{Violation in time domain}\label{vio.timedomain}

Disturbances by measurements were completely 
neglected when deriving \eref{kubof}, 
as discussed in \sref{sec:wrong}.
On the other hand, we have shown in \ref{sec:1stm} that 
the disturbances on additive observables are $O(\sqrt{N})$
even if measurements are quasiclassical, 
i.e., 
even if they emulate classical ideal measurements as closely as possible.
This result has a great impact on the FDT, as follows.

To measure the temporal equilibrium fluctuation, 
one takes $F(t)=0$
and measures the time correlation of 
$\hat{a}(0) = \hat{A}(0)/\sqrt{N}$ and $\hat{b}(t) = \hat{B}(t)/\sqrt{N}$.
Even if this measurement is quasiclassical,
\begin{equation}
\mbox{disturbances on $\Delta \hat{a}$ and $\Delta \hat{b}$}
= O(\sqrt{N})/\sqrt{N} 
= O(1),
\end{equation}
which does not vanish even in the thermodynamic limit.
Therefore, 
{\em disturbances are significant when measuring fluctuations
however large the system is.}
In fact, 
by fully taking account of the disturbances, we have shown that 
the observed fluctuation is the symmetrized time correlation, as
\eref{eq:corr.ab},
which does not agree with the canonical time correlation:\footnote{
Although they would agree with each other when 
$[ \hat{A}, \hat{H} ] = [ \hat{B}, \hat{H} ] = 0$,
the necessary condition \eref{noncomm_for_kubof} is 
not satisfied in such a case.
}
\begin{equation}
\langle 
\mbox{$\frac{1}{2}$}
\{ \Delta \hat{a}, \Delta \hat{b}(t) \}
\rangle\sub{eq}
\neq
\braket{ \Delta \hat{a} ; \Delta \hat{b}(t) }\sub{eq}.
\end{equation}
Hence, 
the time correlation of the Kubo formula is different from the observed one.

To measure the response function $\Phi_{ba}(t)$ in \eref{LRR}, 
one applies $F(t)$ to the system 
and measures the induced change of $\hat{B}/N$,
i.e., 
$\braket{\hat{B}}_t/N - \braket{\hat{B}}\sub{eq}/N$.
We note that
there is a method of measuring $\Phi_{ba}(t)$
with which disturbances are completely irrelevant,
as explained, e.g., in \S 2 of \cite{Kubo},
because in this method 
measurement is made only {\em once in each run of experiments}.
With this disturbance-irrelevant method, one can 
even use a detector whose measurement error
$\delta B\sub{err}$ is infinitesimal,
which means that its disturbance is much larger than those of quasiclassical 
detectors.
However, such a method is not used in ordinary experiments, 
but rather,  
one will perform {\em multi-time (or continuous) measurements}
to obtain values of $B/N$ at various times.
In such a case, disturbances could be relevant.
Therefore, it is necessary to investigate whether 
the same result is obtained 
as that obtained with the disturbance-irrelevant method.
Let us investigate this point in a typical experiment for 
inspecting the FDT, 
where one measures 
$\hat{B}/N$ (to obtain $\Phi_{ba}$)
using the {\em same quasiclassical detector that is used in the fluctuation measurement.}
For such an experiment, 
assuming that the conditions 
(in the second paragraph of \ref{sec:assmp.eq})
for the QCLT 
are satisfied for sufficiently small $F(t)$,
we have 
\begin{equation}
\mbox{disturbance on } \hat{B}/N = O(\sqrt{N})/N = O(1/\sqrt{N}),
\end{equation}
which is negligible for sufficiently large $N$.
That is, 
{\em disturbances are negligible when measuring response functions
quasiclassically.}
Therefore, for $\Phi_{ba}(t)$ this experiment gives 
the same result as the experiment with the disturbance-irrelevant method.
%



In the derivation of the Kubo formula in \cite{Kubo}, 
a response function was calculated neglecting the disturbance,
while the time correlation was 
obtained just as 
a result of the calculation of the response function.
Hence, according to the above result, 
{\em the formula may be correct 
as a recipe to calculate the response function},\footnote{
For other possible problems of the Kubo formula, 
see, e.g., \cite{SK}.
}
{\em while measured fluctuation is described by a different expression,}
as shown in \Fref{diagram2}.
%
Consequently, 
the FDT is violated as a relation between observed quantities.


\subsection{Violation in frequency domain}\label{sec:vio_omega}

To see the FDT violation more clearly and more concretely, 
we analyze it in the frequency domain in this section.

\subsubsection{Fourier transform and causality.}\label{FTandcausality}

In experiments, it is customary to measure the linear response to an external 
field of a constant frequency.
In such a case, one obtains the Fourier transform 
\begin{equation}
\chi_{ba}(\omega)
=
\int_0^{\infty} \Phi_{ba}(t) 
\ e^{i \omega t} dt,
\label{def:chi}
\end{equation}
which is called a (generalized) {\em admittance}.
The dynamical magnetic susceptibility and electrical conductivity are 
examples of the admittance.
Here, the lower limit of integration is $0$ because of the causality,
i.e., 
\begin{equation}
\Phi_{ba}(t) = 0$ for $t<0.
\label{eq:causality}\end{equation}
This is important because, e.g., it leads to the dispersion 
relation \cite{KTH,Zubarev,SY2010}, which is a universal relation between the real and 
imaginary parts of $\chi_{ba}(\omega)$.
It plays an important role in experimental analyses
because, e.g., one can estimate the real part by measuring the imaginary part,
and vice versa.

In contrast to $\Phi_{ba}(t)$, 
the correlation $\tilde{\Xi}_{ba}(t)$, 
\eref{eq:corr.ab}, does not vanish for $t<0$. 
It might thus look natural to consider its full Fourier transform,
\begin{equation}
\tilde{S}_{ba}(\omega)
\equiv
\int_{-\infty}^{\infty} \tilde{\Xi}_{ba}(t)\ e^{i \omega t} dt,
\label{def:tildeSomega}\end{equation}
where the lower limit of integration is extended to $-\infty$.
This quantity can be measured directly in experiments, and hence will  
be used when expressing our results in later sections.
However, when investigating the validity of the FDT,
it is not an appropriate quantity to compare with $\chi_{ba}(\omega)$ 
because then the FDT would be partially violated even for classical systems,
as shown in \ref{app:apvio}.
Since such a superficial violation is not interesting to us,
we compare $\chi_{ba}(\omega)$ with 
\begin{eqnarray}
S_{ba}(\omega)
&\equiv
\int_{0}^{\infty} \Xi_{ba}(t) \ e^{i \omega t} dt,
\label{def:Somega}
\end{eqnarray}
where the lower limit of integration is taken $0$ as in \eref{def:chi}.
Then, the FDT states
\begin{equation}
\chi_{ba}(\omega)
\stackrel{\mbox{?}}{=}
\beta S_{ba}(\omega)
\quad \mbox{ for all } \omega.
\label{FDT:omega}
\end{equation}
Let us investigate whether it would hold
as a relation between observed quantities.
In doing so, 
{\em we assume that the Kubo formula is 
a correct recipe to calculate the response function}, 
as discussed in \ref{vio.timedomain}.



\subsubsection{Symmetric and antisymmetric parts.}\label{sec:sandas}

The admittance $\chi_{ba}(\omega)$ 
represents the response of $\hat{B}$ to $Fe^{-i \omega t}$ that 
couples to 
$\hat{C}$ as \eref{eq:Hext}, where $\hat{A} = d\hat{C}/dt$ as \eref{eq:A=dC}.
If we interchange 
$\hat{A}$ and $\hat{B}$,
we obtain $\chi_{ab}(\omega)$, 
which 
represents the response of $\hat{A}$ to $Fe^{-i \omega t}$ that 
couples to $\hat{D}$, where $\hat{B} = d\hat{D}/dt$.
If the system has the time-reversal symmetry, 
they satisfy the reciprocal relations
\cite{Onsager1,Onsager2,Takahashi,KTH} 
(see \ref{sec:Onsager} for their validity),
\begin{equation}
\chi_{ba}(\omega) = \epsilon_{a} \epsilon_{b} \chi_{ab}(\omega).
\label{reciprocal_0}
\end{equation}
Here, $\epsilon_{a}$ and $\epsilon_{b}$ are the parity ($=\pm 1$) of 
$\hat{a}$ and $\hat{b}$ under the time reversal.
For example, $\epsilon_{j} = -1$ for the current density $\hat{j}$.
The time-reversal symmetry is broken
when, e.g., an external magnetic field $\bm{h}$ is applied to the system.
In such a case, \eref{reciprocal_0} is generalized as
\begin{equation}
\chi_{ba}(\omega; \bm{h}) 
= \epsilon_{a} \epsilon_{b} \chi_{ab}(\omega; - \bm{h}).
\label{reciprocal_h}
\end{equation}
To make this symmetry manifest, we introduce
\begin{equation}
\chi_{ba}^{\pm}(\omega; \bm{h}) 
\equiv
[\chi_{ba}(\omega; \bm{h}) \pm \chi_{ab}(\omega; \bm{h})]/2,
\label{def:chipm}
\end{equation}
which is called the {\em symmetric} ($+$) and the {\em antisymmetric} ($-$) 
{\em parts}
of the admittance \cite{KTH}.

Henceforth, we denote $\chi_{ba}^{\pm}(\omega; \bm{h})$ simply
by $\chi_{ba}^{\pm}(\omega)$.
According to \eref{reciprocal_0}, 
depending on the sign of $\epsilon_{a} \epsilon_{b}$,
{\em either one of $\chi_{ba}^{\pm}(\omega)$
vanishes for all $\omega$ 
if the system has the time-reversal symmetry} (i.e., if $\bm{h} = \bm{0}$).
In the case of the electrical conductivity tensor $\sigma_{\mu \nu}(\omega)$, 
for example, 
the antisymmetric part vanishes when $\bm{h} = \bm{0}$.

To investigate the FDT, \eref{FDT:omega}, we also introduce 
\begin{equation}
S_{ba}^{\pm}(\omega; \bm{h}) 
\equiv
[S_{ba}(\omega; \bm{h}) \pm S_{ab}(\omega; \bm{h})]/2,
\end{equation}
which is henceforth denoted simply by $S_{ba}^{\pm}(\omega)$. 
%
%
Then, \eref{FDT:omega} is equivalent to 
\begin{equation}
\chi_{ba}^{\pm}(\omega)
\stackrel{\mbox{?}}{=}
\beta S_{ba}^{\pm}(\omega)
\quad \mbox{ for all } \omega.
\label{FDT:pm;omega}
\end{equation}
We furthermore introduce
\begin{equation}
\tilde{S}_{ba}^{\pm}(\omega; \bm{h}) 
\equiv
[\tilde{S}_{ba}(\omega; \bm{h}) \pm \tilde{S}_{ab}(\omega; \bm{h})]/2,
\end{equation}
henceforth denoted by $\tilde{S}_{ba}^{\pm}(\omega)$.
It is easily shown that 
$\tilde{S}_{ba}^+$ is real and $\tilde{S}_{ba}^-$ is pure imaginary, i.e., 
\begin{equation}
%
\tilde{S}_{ba}^+(\omega) = \Re \tilde{S}_{ba}^+(\omega),
\
\tilde{S}_{ba}^-(\omega) = i \Im \tilde{S}_{ba}^-(\omega).
%
%
\label{ReIm_tildeS}
\end{equation}
Relations between $S_{ba}^\pm(\omega)$ and $\tilde{S}_{ba}^\pm(\omega)$
are described in \ref{app:SandStilde}.

\subsubsection{inspection of FDT.}\label{sec:examFDTomega}

Since we assume, 
as mentioned in \ref{FTandcausality}, 
that the Kubo formula is a correct recipe to calculate the response function, 
$\chi_{ba}^{\pm}(\omega)$ is obtained from 
\eref{kubof}, \eref{def:chi} and \eref{def:chipm} 
(see \ref{app:tdlkubo} for 
the order of the limit and the integral).
To compare it with $S_{ba}^{\pm}(\omega)$, 
we note the well-known relation 
(see, e.g., Eq.~(4.8) of \cite{Kubo})
\begin{equation}
I_\beta(\omega)
\int_{-\infty}^{\infty} 
\braket{ \Delta \hat{a} ; \Delta \hat{b}(t) }\sub{eq}
\, e^{i \omega t} dt
=
\beta
\int_{-\infty}^{\infty} 
\langle 
\mbox{$\frac{1}{2}$}
\{ \Delta \hat{a}, \Delta \hat{b}(t) \}
\rangle\sub{eq}
\, e^{i \omega t} dt,
\end{equation}
where
\begin{equation}
I_\beta(\omega)
\equiv 
{\beta \hbar \omega \over 2} 
\coth \left({\beta \hbar \omega \over 2} \right)
\sim
\cases{
1 & $(\hbar \omega \ll \kB T)$
\\
\beta \hbar \omega/2 & $(\hbar \omega \gg \kB T)$
}
\label{Ibeta}
\end{equation}
is the factor often encountered in quantum statistical mechanics.
Since $I_\beta(\omega) \to 1$ as $\hbar \omega/\kB T \to 0$, 
the frequency region $\hbar \omega \ll \kB T$
is sometimes called the {\em classical regime}.
Note, however, that it is completely different from the 
{\em classical limit}, $\hbar \to 0$, as shown below.

Using the above equations
and those in \ref{app:SandStilde}, we find 
\begin{eqnarray}
\Re \chi_{ba}^+(\omega)
&= 
\beta \Re S_{ba}^+(\omega)/I_\beta(\omega),
\label{rel:chi+_S}
\\
\Im \chi_{ba}^+(\omega)
&=
\beta \Im S_{ba}^+(\omega)
+
\beta \int_{-\infty}^{\infty} 
{\mathcal{P} \over \omega' - \omega}
\bigg[ 1-  {1 \over I_\beta(\omega')} \bigg]
\Re \tilde{S}_{ba}^+(\omega')
\, {d\omega' \over 2 \pi},
\label{rel:Imchi+_S}
\\
\Re \chi_{ba}^-(\omega)
&=
\beta \Re S_{ba}^-(\omega)
-
\beta \int_{-\infty}^{\infty} 
{\mathcal{P} \over \omega' - \omega}
\bigg[ 1-  {1 \over I_\beta(\omega')} \bigg]
\Im \tilde{S}_{ba}^-(\omega')
\, {d\omega' \over 2 \pi},
%
\label{rel:chi-_S}
\\
\Im \chi_{ba}^-(\omega)
&= 
\beta \Im S_{ba}^-(\omega)/I_\beta(\omega),
\label{rel:Imchi-_S}
\end{eqnarray}
where $\mathcal{P}$ denotes the principal value.
We can inspect the validity of the FDT \eref{FDT:pm;omega} using these formulas.

First of all, 
as $\hbar \to 0$ the above relations 
yield \eref{FDT:pm;omega}.
Therefore, 
{\em in the classical limit 
the FDT holds completely for both symmetric and antisymmetric parts}.
(As discussed in \ref{app:apvio}, this reasonable 
property would be lost if
we employed $\tilde{S}^\pm$ instead of $S^\pm$
as the fluctuation spectrum.)
%
Actually, however, $\hbar$ is finite in quantum systems,
for which the FDT is partially violated as follows.

For the real symmetric part $\Re \chi_{ba}^+(\omega)$, 
\eref{Ibeta} and \eref{rel:chi+_S} show that 
the FDT holds in the classical regime $\hbar \omega \ll \kB T$,
but it is violated for $\hbar \omega \gtrsim \kB T$.
[The same can be said for the imaginary antisymmetric part $\Im \chi_{ba}^-(\omega)$.]
For the real antisymmetric part $\Re \chi_{ba}^-(\omega)$,
\eref{rel:chi-_S} shows that 
{\em the FDT is violated at all $\omega$}, 
even in the classical regime. 
[The same can be said for the imaginary symmetric part $\Im \chi_{ba}^+(\omega)$.]

The last point can be seen clearly by taking $\omega=0$, 
which is completely in the classical regime.
Then \eref{rel:chi-_S} gives
\begin{equation}
\Re \chi_{ba}^-(0)
=
\beta \Re S_{ba}^-(0)
-
\beta \int_{-\infty}^{\infty} 
{\mathcal{P} \over \omega'}
\bigg[ 1-  {1 \over I_\beta(\omega')} \bigg]
\Im \tilde{S}_{ba}^-(\omega')
\, {d\omega' \over 2 \pi}.
%
%
\label{rel:chi-_S(0)}
\end{equation}
Since the last integral does not vanish in general 
for the systems for which $\chi_{ba}^-(\omega) \neq 0$,
the FDT is violated even at $\omega=0$.
To understand this result, 
note that there are two ways to reach the classical regime, $\hbar \omega \ll k_B T$. One is to take the classical limit $\hbar \to 0$, 
where the system becomes classical and the violation disappears.  The other is to take $\omega \to 0$ while keeping $\hbar$ constant, where the violation occurs. 
Therefore, {\em the violation of the FDT is a genuine quantum effect that appears in macroscopic scales} (see also \ref{sec:origin}). 

\subsection{Example -- electrical conductivity tensor}\label{sec:example}

As an example, we consider the electrical conductivity tensor 
$\sigma_{\mu \nu} (\omega)$ 
($\mu, \nu = x,y$\footnote{
Although $\mu, \nu$ correspond to $j_\mu, j_\nu$, respectively, 
according to our notation of $\chi_{ba}(\omega)$,
we write $\sigma_{j_\mu j_\nu}$ as $\sigma_{\mu \nu}$
to indicate that it is a tensor.
})
in a magnetic field $\bm{h} =(0,0,h)$.
Since we assume that the Kubo formula is a correct recipe, 
the observed conductivity (admittance) is given by
\begin{equation}
\sigma_{\mu \nu} (\omega)
=
\int_0^{\infty} 
\langle 
\hat{j}_\nu; \hat{j}_\mu(t)
\rangle\sub{eq}
\ e^{i \omega t} dt,
\label{eq:sigmaomega}
\end{equation}
where $\hat{j}_\nu$ denotes 
the $\nu$ component 
of the total current divided by $\sqrt{N}$. 
We compare it with the observed spectrum intensity of the fluctuation,
\begin{equation}
S_{\mu \nu}(\omega)
=
\int_0^{\infty} 
\langle 
\mbox{$\frac{1}{2}$}
\{ \hat{j}_\nu, \hat{j}_\mu(t) \}
\rangle\sub{eq}
\ e^{i \omega t} dt,
\label{eq:Somega}
\end{equation}
where we have taken the lower limit of integration $0$,
as discussed in \ref{FTandcausality}.

We consider a system that 
is invariant under rotation by $\pi/2$ about the $z$ axis. 
Then the obvious symmetries 
$\sigma_{x y} = - \sigma_{y x}$ and so on yield
\begin{eqnarray}
\mbox{symmetric parts: }
\sigma_{xx}^+ = \sigma_{yy}^+ = \sigma_{xx} = \sigma_{yy},
\
\sigma_{xy}^+ = \sigma_{yx}^+ = 0,
\\
\mbox{antisymmetric parts: }
\sigma_{xx}^- = \sigma_{yy}^- = 0,
\
\sigma_{xy}^- = - \sigma_{yx}^- = \sigma_{xy} = - \sigma_{yx},
\end{eqnarray}
and similarly for $S_{\mu \nu}^\pm$ and $\tilde{S}_{\mu \nu}^\pm$.
Hence, the symmetric and antisymmetric parts 
of $\sigma_{\mu \nu}$ are
the diagonal and the off-diagonal (Hall) conductivity, 
$\sigma_{x x}$ and $\sigma_{x y}$, respectively.

At $\omega=0$, \eref{def:chi} and \eref{def:Somega}
show that 
$\sigma_{\mu \nu} (0)$ and $S_{\mu \nu} (0)$ 
are real because $\Phi_{ba}(t)$ and $\Xi_{ba}(t)$ are real.
Hence, 
one is usually more interested in the real part, 
$\Re \sigma_{x x}$ and $\Re \sigma_{x y}$,
although $\Im \sigma_{x x}$ and $\Im \sigma_{x y}$ are finite for 
$\omega \neq 0$.
Let us therefore study $\Re \sigma_{x x}$ and $\Re \sigma_{x y}$.

For the real symmetric part,
$\Re \sigma^+_{x x} = \Re \sigma_{x x}$,
\eref{rel:chi+_S} gives
\begin{eqnarray}
\Re \sigma_{xx}(\omega)
&= 
\beta \Re S_{xx}(\omega)/I_\beta(\omega)
\label{FDT:sigmaxx}
\\
&=
\cases{
\beta \Re S_{xx}(\omega)
& $(\hbar \omega \ll \kB T)$, 
\\
\displaystyle
(2 / \hbar \omega)
\Re S_{xx}(\omega)
& $(\hbar \omega \gg \kB T)$. 
}
\end{eqnarray}
%
Therefore, 
the FDT holds in the classical regime $\hbar \omega \ll \kB T$,
whereas it is violated for $\hbar \omega \gtrsim \kB T$.
Interestingly, 
\eref{FDT:sigmaxx} gives
$
I_\beta(\omega) \Re \sigma_{xx}(\omega) 
= 
\beta \Re S_{xx}(\omega)
$,
the integral of which over $\omega$
(i.e., the relation for the $t=0$ components)
coincides with Eq.~(4.8) of Callen and Welton \cite{CW}
(who did not take account of disturbances by measurements), 
but not with the results of Nakano \cite{Nakano}, Kubo \cite{Kubo}, 
or Nyquist \cite{Nyquist}.
This is because, as discussed in \ref{sec:2ndm},
one can forget about disturbances if the symmetrized time 
correlation is employed 
from the beginning as Callen and Welton did.

Regarding the real antisymmetric part,
$\Re \sigma^-_{x y} = \Re \sigma_{x y}$,
we compare it with the real part of 
$S^-_{xy}(\omega)
=
S_{xy}(\omega)
=
\int_0^{\infty} 
\langle 
\mbox{$\frac{1}{2}$}
\{ \hat{j}_x, \hat{j}_y(t) \}
\rangle\sub{eq}
\ e^{i \omega t} dt
$.
%
%
For a system invariant under the rotation by $\pi/2$ about the $z$ axis,
they become finite only when a magnetic field $\bm{h}$ is applied.
According to \eref{rel:chi-_S},
they are related by
\begin{equation}
\Re \sigma_{xy}(\omega)
=
\beta \Re S_{xy}(\omega)
-
\beta \int_{-\infty}^{\infty} 
{\mathcal{P} \over \omega - \omega'}
\bigg[ 1-  {1 \over I_\beta(\omega')} \bigg]
\Im \tilde{S}_{xy}(\omega')
\, {d\omega' \over 2 \pi}.
%
%
\end{equation}
Therefore, when $\bm{h} \neq 0$, 
the FDT is {\em violated at all $\omega$}, 
including the classical regime.
Even at $\omega=0$, 
for which $\sigma_{xy}(0)$ and $S_{xy}(0)$ are real,
it is violated because  
\begin{equation}
\sigma_{xy}(0)
=
\beta S_{xy}(0)
-
\beta \int_{-\infty}^{\infty} 
{\mathcal{P} \over \omega'}
\bigg[ 1-  {1 \over I_\beta(\omega')} \bigg]
\Im \tilde{S}_{xy}(\omega')
\, {d\omega' \over 2 \pi},
%
%
\end{equation}
where the last integral does not vanish when $\sigma_{xy} \neq 0$ 
since its integrand is the product of 
the odd function $1/\omega'$, 
the even one $1- 1/I_\beta(\omega')$,
and the odd one $\tilde{S}_{xy}(\omega')$.
This violation should be confirmed experimentally, 
by measuring 
$\sigma_{xy}(0)$ and $S_{xy}(0)$ independently.

Note that $\sigma_{xy}$ is not related to dissipation directly,
because the power supplied by an electric field 
$\bm{E} = (E,0,0)$ is
given by the diagonal conductivity as 
$\bm{E}\cdot\bm{j} = \sigma_{xx} E^2$.
However, 
$\sigma_{xy}$ is surely a property of a nonequilibrium state
because when measuring $\sigma_{xy}$ one must apply $\bm{E}$,
which drives the system into a nonequilibrium state and dissipation occurs
(except in the extreme case where the quantum Hall effect occurs, 
for which $\sigma_{xx}=\sigma_{yy}=0$ 
and hence dissipation is absent).

\subsection{Experiments on violation}

\subsubsection{Notice.}

For $\Re \sigma_{xx}(\omega)$ at $\hbar \omega \ll \kB T$,
all the previous theoretical results on the FDT for quantum systems \cite{CW,Nakano,Kubo} 
and the present one
agree with each other and with the classical results \cite{Nyquist,Takahashi,Green}.
This fact suggests that 
the FDT is relatively insensitive to the choice of measuring apparatuses
for the real symmetric part in the classical regime $\hbar \omega \ll k_B T$.
%
%
In fact, 
many experimental evidences for this case 
have been reported that support the FDT 
(as mentioned in \sref{sec:intro}),
in agreement with these theoretical results including ours,
although conventional measuring apparatuses were used in these experiments,
without considering whether they are quasiclassical.

By contrast, greater care is necessary when inspecting
the FDT for other cases, e.g., 
for the real symmetric part at higher frequencies
and for the real antisymmetric part.
In these cases, 
our results predict the violation.
To confirm this prediction experimentally, 
measurements should be quasiclassical 
because otherwise 
disturbances by measurements would be larger and consequently the 
FDT would look 
violated {\em more greatly},
and one could not tell whether the FDT is really violated.
To avoid such a superficial violation,
quasiclassical measurements should be made, which emulate
classical ideal measurements.

%
%

Note that 
conventional measurements are not necessarily quasiclassical. 
When measuring electromagnetic fields, 
for example, 
Glauber showed that conventional photodetectors,
such as photodiodes and photomultipliers, 
 destroy the state by absorbing photons,
and consequently they 
cannot measure, e.g., the zero-point fluctuation 
\cite{Glauber,Gardiner,MW,WM}.
Such detectors are not quasiclassical,
and hence are not appropriate 
for inspecting the validity of the FDT.

Since quasiclassical measurements are general measurements 
that satisfy the conditions of \ref{sec:assmp.ms},
there are various ways to realize them.
For example, quasiclassical measurements
may be possible by using the heterodyning technique 
(see \ref{sec:vio_high_f}) 
or the quantum non-demolition detectors 
such as those proposed in 
\cite{PSJmeeting,IQEC,QNDpra,QND1,QND2,ISQM}.

\subsubsection{Violation at high frequencies.}\label{sec:vio_high_f}

Koch \etal \cite{Koch} reported
a pioneering experiment 
on the real symmetric parts 
$\Re \sigma_{xx}(\omega)$ and $\Re S_{xx}(\omega)$
($=\tilde{S}_{xx}(\omega)/2$ according to 
\eref{ReIm_tildeS} and \eref{ReS+_tS+}).
They used the heterodyning technique, which 
is closer to quasiclassical
than 
conventional detectors,
because it does not destruct states by absorbing quanta.
In fact, a theoretical analysis \cite{Milburn}\footnote{
Milburn \cite{Milburn}. 
In quantum optics the outgoing modes at different times commute
with each other.
Hence, the last term in Eq,~(3.33) of this reference equals
the corresponding symmetrized time correlation.
}
of the heterodyning technique
shows that the symmetrized time correlation is obtained, 
in agreement with our result on quasiclassical measurement.

The system studied by Koch \etal is 
a resistivity-shunted Josephson Junction,
for which $\Re \sigma_{xx}(\omega)$ is nearly independent of $\omega$
for $\omega/2 \pi \lesssim 2 \times 10^{12}$Hz. 
Hence, if the FDT held $\Re S_{xx}(\omega)$ would also 
be independent of $\omega$.
However, they found that $\Re S_{xx}(\omega)$ increases 
with increasing $\omega$ 
for $\omega/2 \pi \gtrsim 3 \times 10^{10}$Hz.
This shows that the FDT is violated at such high frequencies, 
whereas it holds at lower frequencies,
in agreement with \eref{FDT:sigmaxx}.\footnote{\label{fn:agreement}
When a conventional detector was used, they observed that 
$S_{xx}(\omega)$ {\em decreases}
with increasing $\omega$.
This agrees {\em not} with Callen and Welton \cite{CW}, Nakano \cite{Nakano}, or
Kubo \cite{Kubo}, 
but with Nyquist \cite{Nyquist},
who introduced a quantum effect intuitively into his classical theory.
}

\subsubsection{Violation at low frequencies.}

For the real symmetric part, 
such a high frequency as in \cite{Koch} is necessary to observe the violation
because the FDT is not violated in the classical regime.
This seems a reason why the FDT violation was not found 
in earlier experiments such as the pioneering 
experiment by Johnson \cite{Johnson}.

For the real antisymmetric part, 
by contrast, we have shown that 
the FDT is violated at all $\omega$, 
even in the classical regime including $\omega=0$.
To the authors' knowledge, no experiments have been 
reported which inspected the validity of the FDT for 
the real antisymmetric part.
This might be because,
for systems with the time-reversal symmetry,
the antisymmetric part vanishes if the symmetric part is finite,
as discussed in \ref{sec:sandas}.
In the case of the electrical conductivity tensor $\sigma_{\mu \nu}(\omega)$, 
for example, 
the antisymmetric part vanishes when a magnetic field $\bm{h} = \bm{0}$
(for systems invariant under rotation by $\pi/2$ about the $z$ axis),
as discussed in \ref{sec:example}.
It is therefore expected that 
the violation at low frequencies will be 
observed if one measure 
$\sigma^-_{x y} = \sigma_{x y}$
and
$S^-_{xy}
=
S_{xy}
$
independently in the presence of $\bm{h}$.

\section{Discussions}\label{sec:discussion}

\subsection{Why quantum effects survive on the macroscopic scale?}
\label{sec:origin}

We have shown that the violation of the FDT is a genuine quantum effect.
On the other hand, the FDT relates the response and fluctuation 
of macrovariables.
One might question why a 
quantum effect survives on the macroscopic scale.

This question might be based on the following argument.
Consider two additive operators, 
\begin{equation}
\hat{A} = \sum_{\bm{r}} \hat{\xi}(\bm{r}), 
\
\hat{B} = \sum_{\bm{r}} \hat{\zeta}(\bm{r}).
\end{equation}
Their {\em densities} tend to commute as $N \to \infty$:
\begin{eqnarray}
&
[ \hat{A}/N, \hat{B}/N ]
= {1 \over N^2} \sum_{\bm{r}} [\hat{\xi}(\bm{r}), \hat{\zeta}(\bm{r})]
= {1 \over N^2} \ O(N) \to 0.
\label{comrel.densities}
\end{eqnarray}
This equation looks as if it showed that 
the system would behave like a classical system
for sufficiently large $N$
when one looks at densities of additive observables.

Such a naive argument is false.
For example, 
one can induce magnetization $\bm{M}/N$, which is the density of 
magnetic moments,
by applying a static magnetic field to a material
even if contribution from spins are absent.
However, 
according to the Bohr-van Leeuwen theorem,
$\bm{M}/N$ coming from orbital motions of classical particles vanishes in any equilibrium states.
Therefore, the 
magnetism by orbital motions is a quantum effect that survives on the macroscopic scale.
Although $\hbar$ is small, 
its effect on each particle is significant,
and a collection of Avogadro's number of particles
yields a quantum effect on the macroscopic scale,
because a small number ($\propto \hbar$) 
times a large number ($\propto$ Avogadro's number) is an ordinary number.

Furthermore, \eref{comrel.densities} is not appropriate for
fluctuations, because fluctuations are $O(\sqrt{N})$.
If we scale $\Delta \hat{A}, \Delta \hat{B}$ correctly, we find
\begin{eqnarray}
&
[ \Delta \hat{A}/\sqrt{N}, \Delta \hat{B}/\sqrt{N} ]
= [ \Delta \hat{a}, \Delta \hat{b} ]
= O(1),
\label{comm.fluc}\end{eqnarray}
which does not vanish even in the thermodynamic limit.
This clearly shows that 
{\em disturbances are significant when measuring fluctuations
however large the system is.}

\subsection{Quantum violation of Onsager's regression hypothesis}
\label{sec:Onsager}

In his famous papers \cite{Onsager1,Onsager2}, 
Onsager assumed 
``the average regression of equilibrium fluctuations will 
obey the same laws as the corresponding macroscopic irreversible processes."
Under this hypothesis, 
called the `regression hypothesis,'
he derived the reciprocal relations for classical systems.
For classical systems, 
this hypothesis and the FDT 
were proved by Takahashi
microscopically from Newtonian mechanics \cite{Takahashi}.

For quantum systems, contradictory opinions have been 
claimed.
Assuming that the symmetrized time correlation 
is the equilibrium fluctuation in the hypothesis, 
Kubo and Yokota \cite{KY}, 
Talkner \cite{Talkner}, and Ford and O'Connel \cite{Ford}
 pointed out that the hypothesis is inconsistent with the Kubo formula.
On the other hand, 
Nakajima showed that the inconsistency can be removed if
a local equilibrium state is assumed for the state during fluctuation
\cite{Nakajima56}.
His idea was incorporated in \cite{KYN}, where 
a quantum-mechanical formula for responses to 
`non-mechanical forces' (such as the temperature difference) was derived.

These contradicting opinions originated from
different assumptions.
Unfortunately, as in the case of the FDT,
it was hard to examine the assumptions 
at the time of the above pioneering works because 
neither quantum measurement theory 
\cite{Glauber,Gardiner,MW,WM,KSreview} 
nor theory of macroscopic quantum systems 
\cite{goderis,matsui,Mthesis,Lieb,Nachtergaele,LRB_cont}
was developed enough. 
With the help of development of these theories in the last few decades,
we have {\em proved} that the 
symmetrized time correlation is {\em always} obtained 
when the time correlation of macrovariables is measured 
by quasiclassical measurements.
That is, 
when the regression of equilibrium fluctuations
is really measured the symmetrized time correlation
is obtained. This justifies the above-mentioned assumption by 
Kubo and Yokota \cite{KY}, 
Talkner \cite{Talkner}, and Ford and O'Connel \cite{Ford}.
Therefore, 
the regression hypothesis cannot be valid in quantum systems as 
observed regressions. 

Then, one might wonder if the reciprocal relations hold in quantum systems
because Onsager derived them from the regression hypothesis \cite{Onsager1,Onsager2}.
Fortunately, however, 
the reciprocal relations hold if
the Kubo formula is a correct recipe to calculate the response function,
because they can be derived from the Kubo formula
without the regression hypothesis \cite{Kubo}.

Furthermore, 
the quantum-mechanical formula for responses to non-mechanical forces in \cite{KYN} may also be justified by regarding 
the local equilibrium state in the theory 
not as the state that is observed during fluctuation 
(which is analyzed in \ref{sec:1stm} and \ref{sec:relax})
but as 
the local equilibrium state that would be realized 
as an initial state under appropriate constraints.

\subsection{Relaxation of squeezed equilibrium state}\label{sec:relax}

\begin{figure}[tp]
\begin{center}
\includegraphics[width=0.6\textwidth]{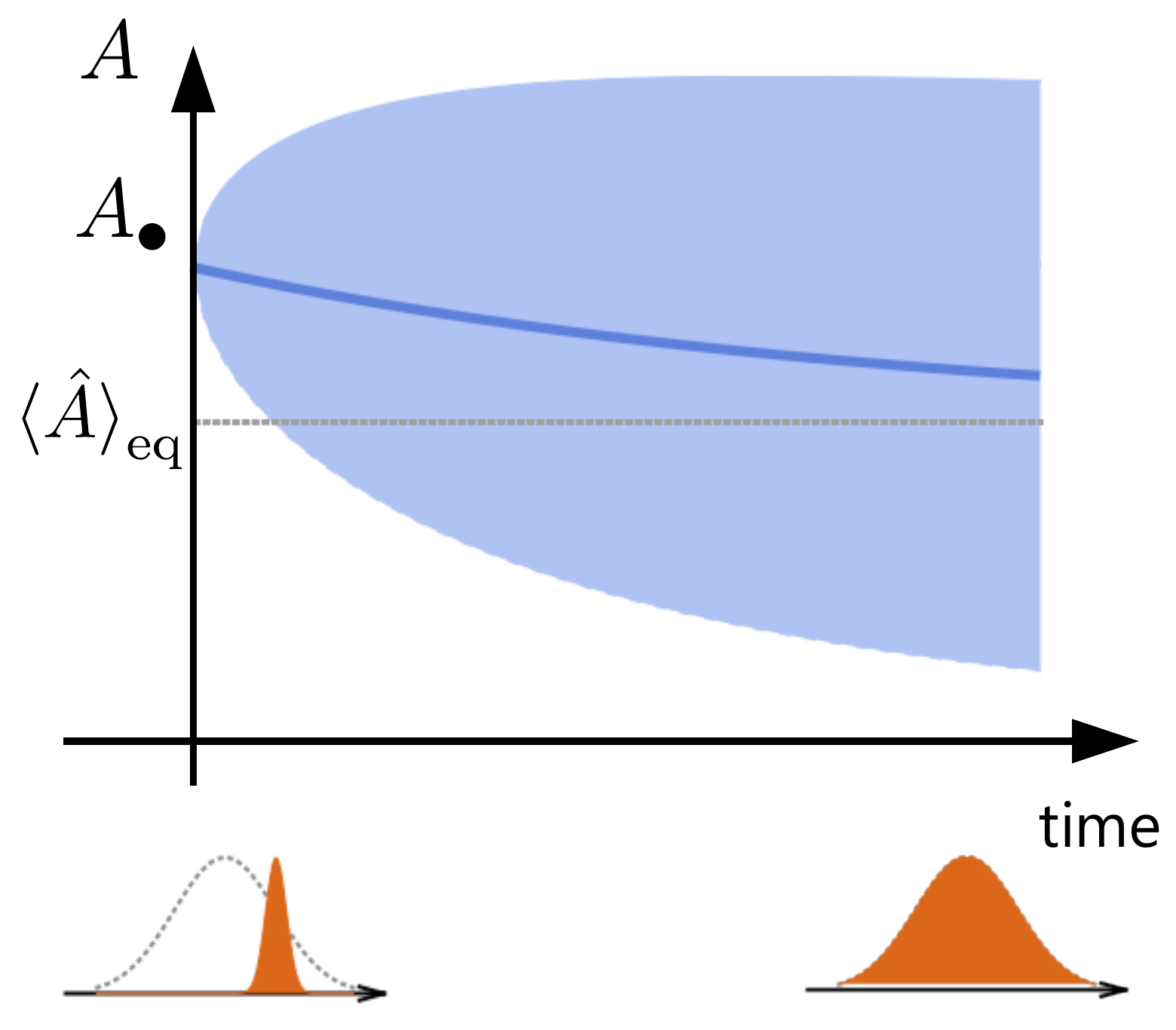}
\end{center}
\caption{Schematic plots of the distribution of an additive observable $A$ 
during the interval between the first and the second measurements.
}
\label{sqeq_relax}
\end{figure}

It is seen in \ref{sec:1stm} that 
the post-measurement state 
$| \beta; a_\bullet \ket$ 
of the first measurement 
is a squeezed equilibrium state,
which is macroscopically identical to $| \beta \ket$
but is squeezed along $\hat{a}$, as shown in \Fref{post_m_state_a}.
During the interval between the first and the second measurements,
the system evolves freely.
In this interval, the expectation value
and the variance of an additive operator $\hat{B} = \hat{b} \sqrt{N}$
is calculated, for the Gaussian $f$, as
\begin{eqnarray}
\fl
\braket{\hat{b}(t)}_{a_\bullet}
=
\braket{\hat{b}}\sub{eq}
+
{
\langle 
\frac{1}{2} \{ \Delta \hat{a}, \Delta \hat{b}(t) \}
\rangle\sub{eq}
%
%
\over
\delta a\sub{eq}^2 + \delta a\sub{err}^2
}
\Delta a_\bullet
\qquad \mbox{(for Gaussian $f$)},
\\
\fl
\langle (\hat{b}(t) - \langle \hat{b}(t) \rangle_{a_\bullet})^2 \rangle_{a_\bullet}
=
\delta b\sub{eq}^2
-
{
\langle 
\frac{1}{2} \{ \Delta \hat{a}, \Delta \hat{b}(t) \}
\rangle\sub{eq}^2
\over
\delta a\sub{eq}^2 + \delta a\sub{err}^2
}
+
{
\braket{\mbox{$\frac{1}{2i}$} [ \hat{a}, \hat{b}(t) ]}\sub{eq}^2
\over 
\delta a\sub{err}^2
}
\quad \mbox{(for Gaussian $f$)}.
\end{eqnarray}
Hence, as shown in \Fref{sqeq_relax}, 
they evolve with increasing $t$,
and relax to the original values,
$\braket{\hat{b}}\sub{eq}$ and $\delta b\sub{eq}^2$,
if the system has the ``mixing property" in the sense that
\begin{equation}
\lim_{t \to \infty}
\lim_{N \propto V \to \infty} 
\langle 
\mbox{$\frac{1}{2}$} \{ \Delta \hat{a}, \Delta \hat{b}(t) \}
\rangle\sub{eq}
= 0
\ \mbox{ and } \
\lim_{t \to \infty}
\lim_{N \propto V \to \infty} 
\braket{\mbox{$\frac{1}{2i}$} [ \hat{a}, \hat{b}(t) ]}\sub{eq}
= 0
\end{equation}
are both satisfied.
After the relaxation, 
one cannot distinguish 
$| \beta; a_\bullet \ket$ from $| \beta \ket$
by macroscopic observations,
i.e., the system `thermalizes.' 

In short, 
the squeezed equilibrium state $| \beta; a_\bullet \ket$ 
is a time-evolving state, 
in which macrovariables fluctuate and relax, 
unlike the thermal pure quantum state $| \beta \ket$ or 
the Gibbs state $e^{-\beta \hat{H}}/Z$.
Such a state should be realized during quasiclassical measurements of temporal equilibrium fluctuations.

\subsection{Measurement with lower time resolution}

In real experiments, the time resolution of 
a detector is finite.
Let us consider how 
\eref{rel:chi+_S} and \eref{rel:chi-_S}
are modified when
$\chi_{ba}^\pm(\omega)$ and $S_{ba}^\pm(\omega)$
are measured with such a detector.

For simplicity, we model a detector of finite time resolution
as a combination of a low-pass filter and 
a (quasiclassical) detector of infinitesimal time resolution.
For the low-pass filter, 
we assume an ideal one, whose frequency response 
$w(\omega)$ is a smooth real function that satisfies
\begin{equation}
w(0)=1, \
w(-\omega) = w(\omega), \
|w(\omega)| \ll 1 \mbox{ for } \omega \gg \Omega. 
\end{equation}
Here, $\Omega >0$ is the cutoff frequency of the filter.

When $\hat{B}$ is measured with such a detector of finite time resolution,
what are obtained as 
$\chi^\pm(\omega)$ and $S^\pm(\omega)$
(the subscript $ba$ is omitted here)  
are respectively
\begin{eqnarray}
\chi_w^\pm(\omega) \equiv w(\omega) \chi^\pm(\omega),
\\
S_w^\pm(\omega) \equiv [w(\omega)]^2 S^\pm(\omega).
\end{eqnarray}
From \eref{rel:chi+_S} and \eref{rel:chi-_S}, 
they are related by
\begin{eqnarray}
\fl
\Re \chi_w^+(\omega)
&= 
\beta \Re S_w^+(\omega)/w(\omega) I_\beta(\omega),
\label{rel:chiw+_S}
\\
\fl
\Re \chi_w^-(\omega)
&=
\beta \Re S_w^-(\omega)/w(\omega)
-
\beta w(\omega) \int_{-\infty}^{\infty} 
{\mathcal{P} \over \omega' - \omega}
\bigg[ 1-  {1 \over I_\beta(\omega')} \bigg]
\Im \tilde{S}^-(\omega')
\, {d\omega' \over 2 \pi}.
%
%
\label{rel:chiw-_S}
\end{eqnarray}
Note here that we consider the case where $\tilde{S}^-(\omega)$ 
in the last term 
is not measured because 
it is sufficient for confirming the FDT violation 
to measure $\chi^\pm(\omega)$ and $S^\pm(\omega)$.
The value of the last term is determined by the physical properties of the
system.

Therefore, 
for the symmetric part 
the FDT violation is detectable only when 
$\hbar \Omega \gtrsim \kB T$ and 
$\hbar \omega \gtrsim \kB T$.
By contrast, 
for the antisymmetric part
the FDT violation is detectable 
for all $\omega$ such that $\omega \lesssim \Omega$
even if $\hbar \Omega \ll \kB T$.
In particular, for $\omega=0$
the same result as \eref{rel:chi-_S(0)} is obtained.
Therefore, {\em the FDT violation for the 
antisymmetric part is fully detectable
by a quasiclassical detector even if its time resolution is poor.}


\subsection{Related works}

To investigate the validity of the FDT, 
we have studied equilibrium temporal fluctuations 
of macrovariables in 
quantum systems.
We assume that 
the systems have a macroscopic degrees of freedom, 
and, accordingly, the QCLT 
is applicable.
To exclude a superficial violation which can be induced by strong
disturbances (backactions) by detectors (measuring apparatuses), 
we have assumed quasiclassical detectors 
that emulate classical ideal measurements as closely as 
possible,
rather than ``violent" detectors such as those perform projection measurements.
{\em These two factors, macroscopic degrees of freedom and 
quasiclassical detectors, have enabled us to derive the universal 
results,} which are independent of any details of 
the physical systems and detectors.

Measurements of fluctuations have also been studied by many other works
for various physical systems and in various viewpoints.
We briefly summarize some of such works.

Since the pioneering work by Glauber \cite{Glauber}, 
both theories and experiments on fluctuation measurements 
have been most developed in quantum optics, 
as described in textbooks \cite{MW,WM}.
There, 
although equilibrium states of macroscopic systems are sometimes studied, 
more interest is devoted to measurement and {\em control} of 
photons in a {\em small number} of modes {\em far from} equilibrium
(such as photons emitted from a laser).

Measurements of fluctuations in condensed matter have recently been studied 
intensively.
For measurements of the ``full-counting statistics"
\cite{oNK,oEHM,oHC},  
which is an electron analog of
the photon-counting statistics of quantum optics \cite{Gardiner,MW,WM}, 
explicit results are obtained mostly about small 
open systems which are connected to reservoirs, 
such as mesoscopic systems and systems with a small degrees of freedom.
It is interesting to explore whether a universal result
could be extracted with increasing the degrees of freedom toward 
macroscopic systems.
Measurements are also studied for the fluctuation theorem (FT) in quantum 
systems \cite{oEHM,oCHT}.
We note that the FT is different from the Kubo formula in 
several aspects,
although some textbooks state that the Kubo formula 
can be derived from the FT.
For example, 
the admittances at finite $\omega$ 
can easily be treated by the Kubo
formula, whereas it seems not so easy to treat them by the FT for quantum 
systems.
Furthermore, the FT cannot treat 
the admittances that are not directly related to dissipations, 
such as $\Im \sigma_{xx}$ and $\sigma_{xy}$.
Moreover, 
the FT focuses on dissipations {\em caused by reservoirs}.
By contrast, 
the Kubo formula focuses on dissipations 
{\em in the bulk of the system},
even when a current is induced by reservoirs, 
because the authors of the pioneering works 
\cite{Green,Takahashi,CW,Nakano,Kubo}
including Kubo were interested in 
the fundamental question of how dissipation emerges from 
non-dissipative microscopic dynamics.
(For this reason, the ``mixing property" \eref{mixing_for_kubof} is necessary
for the Kubo formula).
For these essential differences, 
it is not clear how the FDT violation is related to the FT.
More recently, measurements of work are discussed in \cite{oSG,oS,oTH}.
It will be interesting to examine whether disturbances by measurements
could cause violation of some fundamental relations on work.

Note that some of these studies \cite{oEHM,oCHT,oSG} assumed the two-time {\em projection} measurements, whereas we have assumed the two-time (or more-time) 
{\em quasiclassical} measurements.
When a quasiclassical measurement is made 
on {\em any} equilibrium state, 
we have shown in \ref{sec:1stm} that 
the post-measurement state of the 
first measurement is a squeezed equilibrium state, 
which is macroscopically identical to the pre-measurement equilibrium state.
By contrast, 
when a {\em projection} measurement is made 
on an equilibrium state of a certain class of systems,
it is recently shown that 
the post-measurement state becomes a quite anomalous state, far from equilibrium \cite{qit1,qit2,TS2017}.
Therefore, the strong disturbances of projection measurements would lead to a superficial violation of the FDT, which is greater than the violation 
observed by quasiclassical measurements.
This fact shows clearly that,
when examining fundamental relations for macroscopic systems, 
projection measurements are too violent and 
quasiclassical measurements are necessary.

\section{Summary}\label{sec:summary}

We have studied whether the FDT holds as a relation between observed 
quantities in macroscopic quantum systems.
To exclude a superficial 
violation by violent measurements,
we consider the case where measurements are made 
in an ideal way that emulates classical ideal measurements
as closely as possible.
We call such measurements quasiclassical.

Assuming quasiclassical measurements, 
we study 
what is observed when the temporal equilibrium fluctuation is measured.
We have found that the 
symmetrized time correlation is obtained quite generally.
As a result, 
the FDT is violated as a relation between observed quantities.
This is a universal result, which is independent of any details of 
the physical systems and detectors,
as long as the system has macroscopic degrees of freedom and 
the detectors are quasiclassical.
The violation is shown to be 
a genuine quantum effect that survives on a macroscopic scale.

In terms of the (generalized) admittance, 
which is the Fourier transform of the 
response function, the violation is summarized as follows.
For the real symmetric part and the imaginary antisymmetric part,
the FDT is violated at high frequencies $\hbar \omega \gtrsim k_B T$.
A previous experiment on the diagonal conductivity 
$\Re \sigma_{xx}(\omega)$ reported an evidence.

For the real antisymmetric part and imaginary symmetric part, 
the FDT is violated at all frequencies.
It is violated even at $\omega=0$ for the real antisymmetric part
(while the imaginary symmetric part vanishes at $\omega=0$).
To the authors' knowledge, no experiment has been reported
that inspected the FDT in such a case.
The violation should be confirmed experimentally by measuring independently the admittance and the time correlation for the case of, e.g., the Hall conductivity and the corresponding current-current correlation in the presence of a magnetic field.

In measurement of the temporal equilibrium fluctuation,
two- or more-time measurements should be made in each run of experiment.
Just after the first measurement, 
the post-measurement state 
is a squeezed equilibrium state,
which is macroscopically identical 
to the Gibbs and the thermal pure quantum state
but is squeezed by the measurement.
It is a time-evolving state, 
in which macrovariables fluctuate and relax, 
unlike the Gibbs or the thermal pure quantum state.
Such an interesting state should be 
realized during quasiclassical measurements of 
temporal equilibrium fluctuations.

\ack
We thank 
H. Tasaki, K. Asano, H. Hayakawa, M. Koashi, Y. Oono and N. Shiraishi
for discussions.
This work was supported by The Japan Society
for the Promotion of Science, KAKENHI No. 26287085
and No. 15H05700.

\appendix


\section{Order of various limits and integral in Kubo formula}
\label{app:tdlkubo}

In this appendix, we note a point that is important to perform 
consistent calculations.
Although this point is often disregarded in the literature, 
careless treatments lead to many unphysical results, 
which are often found in the literature.

Substituting \eref{kubof} for $\Phi_{ba}(t)$ in \eref{def:chi}, we have
\begin{equation}
\chi_{ba}(\omega)
=
\int_0^{\infty} 
\lim_{N \propto V \to \infty} 
\beta \braket{ \Delta \hat{a} ; \Delta \hat{b}(t) }\sub{eq}
\ e^{i \omega t} dt.
\label{chi.kubo1}
\end{equation}
Since the thermodynamic limit is taken before the time integral,
this formula is not useful for studying properties of 
$\chi_{ba}(\omega)$.
We therefore rewrite it as follows.

We assume that 
a necessary conditon \eref{remote_for_kubof} for the Kubo formula
is satisfied.
Then, it would be safe to rewrite \eref{chi.kubo1} as
\begin{equation}
\chi_{ba}(\omega)
=
\lim_{\epsilon \searrow 0}
\int_0^{\infty} 
\lim_{N \propto V \to \infty} 
\beta \braket{ \Delta \hat{a} ; \Delta \hat{b}(t) }\sub{eq}
\ e^{i \omega t - \epsilon t} dt.
\label{chi.kubo2}
\end{equation}
For finite $V$, 
$\braket{ \Delta \hat{a} ; \Delta \hat{b}(t) }\sub{eq}$ exhibits
the quantum recurrence, and the recurrence time 
increases with increasing $V$.
Hence, for a given small $\epsilon >0$, 
it is expected that the $V$ dependence of 
$
\braket{ \Delta \hat{a} ; \Delta \hat{b}(t) }\sub{eq}
e^{- \epsilon t}
$
becomes negligible 
for sufficiently large $V$.
Therefore, we may rewrite \eref{chi.kubo2} as
\begin{equation}
\chi_{ba}(\omega)
=
\lim_{\epsilon \searrow 0}
\lim_{N \propto V \to \infty} 
\int_0^{\infty} 
\beta \braket{ \Delta \hat{a} ; \Delta \hat{b}(t) }\sub{eq}
\ e^{i \omega t - \epsilon t} dt.
\label{chi.kubo3}
\end{equation}
Since $V$ is finite in this time integral,
this formula is useful for studying properties of 
$\chi_{ba}(\omega)$.
For example, 
one can express the integral using the energy eigenvalues and eigenstates.
[This is impossible for an infinite system because, e.g., 
the Hamiltonian is ill-defined
(although the local Hamiltonian density is well-defined).]
In such a case, however, 
{\em the limit $\epsilon \searrow 0$ should not 
be taken before the thermodynamic limit.}
Otherwise, unphysical results would be obtained, 
which are, unfortunately, often found in the literature.

We have used \eref{chi.kubo1} and \eref{chi.kubo3} 
interchangeably in \ref{sec:vio_omega},
although the limit symbols and the factor $e^{- \epsilon t}$ were  
not written explicitly.

\section{Superficial violation of FDT in classical systems}
\label{app:apvio}

We have compared $\chi_{ba}$ and $S_{ba}$ as
\eref{rel:chi+_S}-\eref{rel:Imchi-_S}. 
Similar 
relations between $\chi_{ba}$ and $\tilde{S}_{ba}$
(defined by \eref{def:tildeSomega})
were known as formal relations \cite{KTH}:
\begin{eqnarray}
\Re \chi_{ba}^+(\omega)
&= 
\beta \Re \tilde{S}_{ba}^+(\omega)/[2 I_\beta(\omega)],
\label{rel:chi+_tildeS}
\\
\Re \chi_{ba}^-(\omega)
&=
\beta \int_{-\infty}^{\infty} 
{\mathcal{P} \over \omega' - \omega}
\cdot
{1 \over I_\beta(\omega')}
\Im \tilde{S}_{ba}^-(\omega')
\, {d\omega' \over 2 \pi}.
%
%
\label{rel:chi-_tildeS}
\end{eqnarray}
As $\hbar \to 0$ they reduce to
\begin{eqnarray}
\Re \chi_{ba}^+(\omega)
&= 
\beta \Re \tilde{S}_{ba}^+(\omega)/2,
\label{rel:chi+_tildeScl}
\\
\Re \chi_{ba}^-(\omega)
&=
\beta \int_{-\infty}^{\infty} 
{\mathcal{P} \over \omega' - \omega}
\Im \tilde{S}_{ba}^-(\omega')
\, {d\omega' \over 2 \pi}.
%
%
\label{rel:chi-_tildeScl}
\end{eqnarray}
\eref{rel:chi+_tildeScl} shows that 
the FDT,
if it is defined as relations between $\chi_{ba}$ and $\tilde{S}_{ba}$, 
holds for $\Re \chi_{ba}^+(\omega)$ in classical systems,
where the factor $1/2$ can be absorbed in the definition of the 
spectrum intensity $\tilde{S}_{ba}^+(\omega)$.
For $\Re \chi_{ba}^-(\omega)$,
however, 
one would expect the corresponding FDT as 
\begin{equation}
\Re \chi_{ba}^-(\omega)
\stackrel{\mbox{?}}{=}
\beta \Re \tilde{S}_{ba}^-(\omega)/2
=0,
\label{rel:wrong_chi-_tildeScl}
\end{equation}
where we have used \eref{ReIm_tildeS}.
This disagrees with the correct relation \eref{rel:chi-_tildeScl}
whenever $\Re \chi_{ba}^-(\omega) \neq 0$.
Therefore, if one compares 
$\chi_{ba}$ and $\tilde{S}_{ba}$,
the FDT looks as if it 
were violated even in classical systems.

By contrast, if one compares $\chi_{ba}$ and $S_{ba}$
as we did in this paper, {\em the FDT 
holds completely in classical systems,} 
as shown in \ref{sec:examFDTomega}. 
Therefore, we consider the above violation \eref{rel:wrong_chi-_tildeScl}
in classical systems 
just as a superficial violation, which 
comes from the improper comparison.

One might suspect that a pair of $\tilde{S}_{ba}(\omega)$ and 
\begin{equation}
\tilde{\chi}_{ba}(\omega)
\equiv
\int_{-\infty}^{\infty} 
\lim_{N \propto V \to \infty} 
\beta \braket{ \Delta \hat{a} ; \Delta \hat{b}(t) }\sub{eq}
\ e^{i \omega t} dt
\label{tildechi.kubo1}
\end{equation}
would be a better choice for the FDT.
However, 
such $\tilde{\chi}_{ba}(\omega)$ 
{\em disagrees with the observed admittance.}
For example, $\tilde{\sigma}_{x x}(\omega)$ thus defined 
has no imaginary part at any $\omega$, whereas the observed 
admittance does have the imaginary part,   
which represents the phase shift of the response.
[See also discussions following \eref{eq:causality}.]
Therefore, the causality of the response function, 
which determines the lower limit of integration over $t$ as \eref{def:chi},
is very important for getting the correct admittance.

To sum up, 
one has to compare $\chi_{ba}$ and $S_{ba}$, as we did in this paper,
to inspect the FDT appropriately.

%
%

\section{Relations between $S_{ba}^\pm(\omega)$ 
and $\tilde{S}_{ba}^\pm(\omega)$}
\label{app:SandStilde}

Using the convolution theorem and 
\begin{equation}
\int_{-\infty}^{\infty}
\Theta(t)
\ e^{i \omega t} dt
=
\pi \delta(\omega) + i {\mathcal{P} \over \omega},
\end{equation}
we can easily show
\begin{equation}
S_{ba}^\pm(\omega)
=
{1 \over 2} \tilde{S}_{ba}^\pm(\omega)
- i \int_{-\infty}^{\infty} 
{\mathcal{P} \over \omega' - \omega}
\tilde{S}_{ba}^\pm(\omega')
\, {d\omega' \over 2 \pi}.
\end{equation}
This yields
\begin{eqnarray}
\Re S_{ba}^+(\omega)
=
{1 \over 2} \Re \tilde{S}_{ba}^+(\omega),
\label{ReS+_tS+}
\\
\Im S_{ba}^+(\omega)
=
-  \int_{-\infty}^{\infty} 
{\mathcal{P} \over \omega' - \omega}
\Re \tilde{S}_{ba}^+(\omega')
\, {d\omega' \over 2 \pi},
\label{ImS+_tS+}
\\
\Re S_{ba}^-(\omega)
=
\int_{-\infty}^{\infty} 
{\mathcal{P} \over \omega' - \omega}
\Im \tilde{S}_{ba}^-(\omega')
\, {d\omega' \over 2 \pi},
\label{ReS-_tS-}
\\
\Im S_{ba}^-(\omega)
=
{1 \over 2} \Im \tilde{S}_{ba}^-(\omega).
\label{ImS-_tS-}
\end{eqnarray}
These relations have been used when deriving the results of 
\ref{sec:examFDTomega}. 

\end{document}